\documentclass[12pt]{article}

\usepackage[a4paper,margin=1in]{geometry}

\usepackage{setspace}
\usepackage{graphicx}
\usepackage{caption}
\usepackage{subcaption}

\usepackage{amsmath,amssymb}

\usepackage{authblk}

\usepackage[
colorlinks=true,
linkcolor=blue,
citecolor=blue,
urlcolor=blue,
breaklinks=true
]{hyperref}

\usepackage{xurl}

\usepackage[
backend=biber,
style=numeric,
citestyle=numeric-comp,
sortcites=true,
sorting=none,
giveninits=true,
maxbibnames=99,
doi=false,
url=false,
isbn=false,
eprint=false
]{biblatex}

\renewbibmacro*{in:}{}

\DeclareFieldFormat[article]{title}{\mkbibemph{#1}}

\DeclareFieldFormat[article]{journaltitle}{#1}

\DeclareFieldFormat[article]{volume}{#1}

\DeclareFieldFormat{pages}{#1}

\renewbibmacro*{journal+issuetitle}{%
  \usebibmacro{journal}%
  \setunit{\addspace}%
  \printfield{volume}%
  \setunit{\addspace}%
  \printtext[parens]{\printfield{year}}%
  \setunit{\addcomma\space}%
  \printfield{pages}%
  \newunit
}

\renewbibmacro*{note+pages}{%
  \printfield{note}
}

\DeclareFieldFormat{doi}{%
  \href{https://doi.org/#1}
       {\nolinkurl{https://doi.org/#1}}%
}

\renewbibmacro*{doi+eprint+url}{%
  \iffieldundef{doi}
    {}
    {\newunit\newblock\printfield{doi}}
}

\title{\textbf{The Physics of Topological Defects in Glasses}}

\author[a,b]{Arabinda Bera}
\author[c]{Peter Schall}
\author[d]{Timothy W. Sirk}
\author[e]{Vijayakumar Chikkadi}
\author[a,*]{Alessio Zaccone}

\affil[a]{Department of Physics ``A. Pontremoli'', University of Milan, Milan, Italy}
\affil[b]{Institute of Physics, University of Greifswald, Felix-Hausdorff-Stra{\ss}e 6, 17489 Greifswald, Germany}
\affil[c]{Institute of Physics, University of Amsterdam, 1098 XH Amsterdam, The Netherlands}
\affil[d]{Polymers Branch, DEVCOM Army Research Laboratory, Aberdeen Proving Ground, Aberdeen, MD 21005}
\affil[e]{Physics Division, Indian Institute of Science Education and Research Pune, Pune-411008, India}
\affil[*]{Corresponding author: \texttt{alessio.zaccone@unimi.it}}
\date{}

\begin{document}

\maketitle

\begin{abstract}
Topological defects play a central role in the mechanical behavior of crystalline materials, yet their relevance to amorphous solids has only recently begun to emerge. Over the last few years, theoretical, computational, and experimental studies have revealed the presence of well-defined topological invariants in vibrational eigenmodes, non-affine displacement fields, and deformation-induced vector fields of glasses. These defects have been shown to correlate strongly with soft spots, localized plastic rearrangements, yielding, and shear-band formation, suggesting a new perspective on the microscopic origins of plasticity in disordered materials. In this review, we provide a comprehensive overview of recent developments in the rapidly growing field of topological defects in glasses. We discuss the underlying theoretical concepts, including Burgers vectors, non-affine plasticity, vibrational modes, and topological invariants, and review recent numerical and experimental advances. Finally, we assess the current achievements, limitations, and open questions, and discuss future directions toward a unified topological description of plasticity and mechanical failure in amorphous solids.
\end{abstract}


\tableofcontents

\newpage
\section{Introduction and Overview}\label{section1}
Glasses are among the most common materials in nature and technology, yet they remain among the most difficult systems in which to identify the microscopic defects that govern deformation and failure. This broad class of materials encompasses structural glasses, metallic glasses, colloidal glasses, granular packings, foams, and many polymeric systems \cite{debenedetti2001,berthier2011,nicolas2018}. Despite their diverse microscopic constituents and interactions, these systems share the defining characteristic of lacking long-range crystalline order while nevertheless exhibiting rigidity, elasticity, yielding, and plastic flow. Understanding the microscopic origin of these properties remains one of the central challenges of condensed matter physics and materials science. In crystalline solids, mechanical response and plastic deformation can be understood in terms of well-defined defects such as dislocations and disclinations, whose structure, dynamics, and interactions provide a powerful framework for connecting microscopic mechanisms to macroscopic behavior \cite{hirth1982,mura1987}. In amorphous solids, however, the lack of an underlying lattice removes the natural geometrical reference used to define such defects. As a result, identifying the structural entities that govern deformation, relaxation, and failure has remained a longstanding problem and has motivated several decades of theoretical, numerical, and experimental research.

Over the past several decades, numerous theoretical frameworks have been developed to address this challenge, including free-volume theories \cite{cohen1959,spaepen1977}, shear transformation zones \cite{argon1979,falk1998,lin2015,ferrero2021}, Eshelby-like localized rearrangements \cite{eshelby1957,picard2004}, geometric frustration approaches based on locally favored structures \cite{frank1952,tanaka2019}, soft spots identified from low-frequency vibrational modes \cite{widmercooper2006,manning2011}, local yield stress descriptions \cite{patinet2016}, and, more recently, machine-learning-based predictors of plastic activity \cite{schoenholz2016,cubuk2017,cubuk2015,bapst2020,boattini2020}. Early applications of unsupervised learning to amorphous materials employed multiresolution network clustering to detect hidden structural organization across multiple length scales directly from experimental and simulation data, providing one of the first systematic attempts to identify structural heterogeneity and defects in glasses without predefined order parameters \cite{Ronhovde2011,Ronhovde2012,Nussinov2016}. Machine learning approaches based on graph neural networks, autoencoders, and geometric deep learning have considerably expanded these capabilities, enabling increasingly accurate predictions of local structure and plastic activity.

Related developments in the physics of jammed packings revealed that the onset and loss of rigidity in disordered solids are governed by critical scaling laws near the jamming transition (which coincides with the random close packing point \cite{Zaccone2022} or maximally random jammed state \cite{Truskett}), including the vanishing of the shear modulus at the unjamming point \cite{Ohern}. Subsequent theoretical work showed that this loss of rigidity can be quantitatively explained from first principles as a consequence of non-affine particle relaxations, which progressively cancel the affine elastic response and drive the shear modulus to zero at isostaticity at the jamming transition \cite{zaccone2011}. More recently, this framework has been embedded within a complete microscopic theory of the jamming transition derived from a reversible particle-bath Hamiltonian, providing a unified description of both the shear modulus above jamming and the viscosity divergence below jamming \cite{zaccone2025jamming}.

While each of these approaches has provided important insights, a complete and universally accepted description of defects in amorphous materials remains elusive. Recent advances have revealed an intriguing connection between the physics of glasses and concepts traditionally associated with topology and field theory. Rather than seeking defects directly in the particle configuration, one may construct vector fields from mechanical or dynamical observables, such as non-affine displacement fields generated during deformation or eigenvector fields obtained from the vibrational modes of the Hessian matrix. These fields can host singular structures characterized by winding numbers and topological charges, analogous to vortices -- anti-vortices (in space dimension $d=2$), and hedgehog defects ($d=3$) encountered in liquid crystals, magnetic systems, and active matter \cite{kleman2003book,shankar2022}. The emergence of such topological defects (TDs) has opened a new perspective on the long-standing problem of identifying structural signatures of plasticity and mechanical heterogeneity in amorphous solids.

It is worth noting that the idea of combining topology with disorder has precedents outside the physics of structural glasses. In particular, topological superfluids confined in disordered environments exhibit orbital, spin and Weyl glass phases, in which quenched disorder gives rise to topologically non-trivial textures and defects while preserving robust topological characteristics of the underlying fields \cite{Volovik2019}. Although the physical degrees of freedom differ substantially from those of amorphous solids, these studies demonstrate that topology can provide a natural language for describing disordered states across very different condensed-matter systems.

%The purpose of this review is to provide a comprehensive account of the development of topological concepts in the physics of glasses and to place recent advances within the broader historical context of defect theory, elasticity, and amorphous plasticity. We begin with a discussion of classical TDs in crystalline solids and the mathematical framework used to characterize them. We then examine how the absence of crystalline order necessitates alternative approaches to defect identification in glasses, highlighting the major theoretical ideas that have shaped the field. Particular emphasis is placed on non-affine deformation fields, vibrational eigenmodes, and the topological defects that emerge within these vector fields. Finally, we discuss recent theoretical, numerical, and experimental developments, together with open questions and future directions that may help establish a unified defect-based description of amorphous materials.

\subsection{Scope and Organization of this Review}\label{sec1_sub7}

The purpose of this review is to provide a unified overview of recent developments concerning topological defects in amorphous solids and their connection to plasticity, yielding, and mechanical failure. Rather than focusing on a single theoretical framework, we discuss a broad range of approaches in which topological concepts emerge from vibrational eigenmodes, non-affine displacement fields, three-dimensional deformation fields etc. Particular emphasis is placed on understanding how these different descriptions are related and what they reveal about the microscopic origins of plastic deformation in glasses.

The review is organized as follows. In Section~\ref{section1}, we introduce the fundamental concepts of topological defects, Burgers vectors, non-affine motion, and vibrational modes, and discuss their relevance to amorphous materials. Section~\ref{section2} reviews recent theoretical and computational developments, including topological defects in vibrational eigenmodes, non-affine displacement fields, localized plastic rearrangements, shear-band formation, Burgers rings, and three-dimensional hedgehog defects. Section~\ref{section3} discusses the current experimental evidence for topological defects in colloidal glasses and other model amorphous systems. Finally, Section~\ref{section4} summarizes the major advances achieved so far, highlights the limitations of existing approaches, and outlines several open questions and future directions for the field.

\subsection{Defects in crystals}\label{sec1_sub1}
In crystalline materials, plastic deformation and mechanical response are largely governed by the presence and dynamics of topological defects such as dislocations and disclinations. As first recognized independently by Taylor, Orowan, and Polanyi in 1934, slip along crystallographic planes cannot be explained by the simultaneous displacement of entire atomic planes, but instead occurs through the motion of localized line defects known as dislocations \cite{taylor1934,orowan1934,polanyi1934}. This realization laid the foundation for modern defect theory and transformed the understanding of strength, plasticity, and work hardening in crystalline solids.

The mathematical description of defects in solids can be traced to the pioneering work of Volterra, who introduced the concept of singular distortions generated by cutting and reconnecting an elastic body \cite{volterra1907}. Building upon these ideas, Burgers introduced a geometrical characterization of dislocations through the closure failure of a circuit surrounding a defect \cite{burgers1939}. Subsequent developments by Frank, Nabarro, and others established the continuum theory of dislocations and disclinations and clarified their role in crystal plasticity, grain boundaries, and rotational incompatibilities \cite{frank1951,nabarro1947,frank1958}. Comprehensive treatments of these concepts can be found in the classic texts of Hirth and Lothe, Mura, and Kléman and collaborators \cite{hirth1982,mura1987,kleman1977,kleman2008}.

Beyond their importance in conventional crystals, topological defects have also played a central role in understanding orientational order and phase transitions. In particular, the work of Steinhardt, Nelson, and co-workers demonstrated how disclination-like defects and geometric frustration can profoundly influence the structure of condensed matter systems, including liquids, glasses, and quasicrystals \cite{steinhardt1981_1,steinhardt1981_2,steinhardt1983,nelson1983}. These developments highlighted the broader significance of topological concepts across different states of matter and established defects as fundamental objects linking geometry, elasticity, and collective behavior.

A key feature of crystalline defects is that they can be characterized by topological quantities that remain invariant under smooth deformations of the surrounding medium. Among the various defect descriptors, the Burgers vector plays a particularly important role. Defined through the closure failure of a Burgers circuit, it constitutes a vector-valued topological invariant that uniquely characterizes the translational mismatch associated with a dislocation and determines its magnitude and orientation. We therefore begin by reviewing the Burgers-vector construction and its significance in defect theory.

\subsection{Burgers Vector and Topological Characterization}\label{sec1_sub2}
A central advance in the theory of crystalline defects was the realization that dislocations can be characterized by a topological quantity known as the Burgers vector \cite{burgers1939,hirth1982,mura1987}. Consider a closed circuit $\mathcal{L}$ constructed in a perfect crystal by following equivalent lattice directions. In the presence of a dislocation, the circuit fails to close, and the resulting closure failure defines the Burgers vector. Within continuum elasticity, the Burgers vector $\mathbf{b}$ can be expressed, in components, as
\begin{equation}
b_i=-\oint_{\mathcal{L}}du_i
=-\oint_{\mathcal{L}}
\frac{\partial u_i}{\partial x^k}dx^k,
\label{eq:burgers}
\end{equation}
where $u_i$ denotes the displacement field and $\mathcal{L}$ is a closed contour surrounding the defect. Latin indices are used to denote Cartesian components. The value of $|\mathbf{b}|$ is independent of the precise shape of the contour provided that the enclosed topological content remains unchanged. 

\begin{figure*}[t]
\centering
\includegraphics[width=\textwidth]{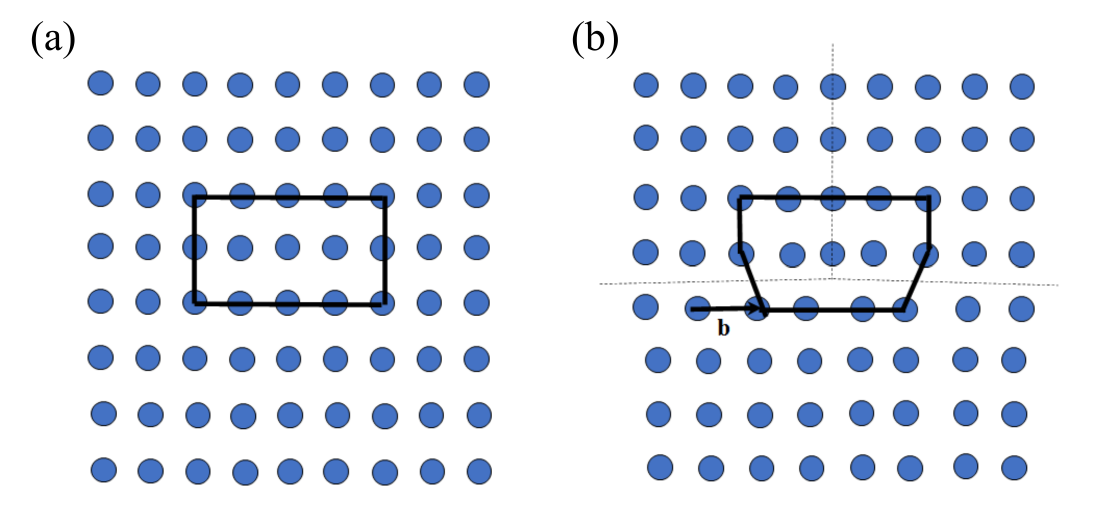}
\caption{Dislocation in crystal and Burgers circuit. (\textbf{a}) Perfect crystal lattice containing no dislocations. A Burgers circuit constructed around the lattice closes exactly, yielding a zero Burgers vector.
(\textbf{b}) Crystal containing an edge dislocation. The same Burgers circuit exhibits a closure failure due to the lattice distortion, defining the Burgers vector \(\mathbf{b}\), which characterizes the magnitude and direction of the defect.}
\label{sec1_fig1}
\end{figure*}

In Fig.~\ref{sec1_fig1}, we schematically compare a perfect crystal with a crystal containing an edge dislocation. While a Burgers circuit closes exactly in the defect-free lattice, it exhibits a closure failure in the presence of a dislocation. This closure failure defines the Burgers vector, the fundamental topological quantity used to characterize the strength and orientation of the defect. The importance of the Burgers vector extends beyond its geometrical interpretation. It provides a conserved topological charge that uniquely characterizes the strength and orientation of a dislocation and establishes a direct connection between microscopic lattice distortions and macroscopic mechanical response \cite{hirth1982,mura1987}. More generally, the Burgers construction exemplifies a powerful principle in defect theory: defects can often be identified and classified through topological quantities that remain invariant under smooth deformations of the surrounding medium \cite{volterra1907,kleman1977,kleman2008}.

This topological viewpoint has proven remarkably successful in crystalline solids, where the existence of an underlying lattice provides a natural reference frame for defining displacement fields and closure circuits. In amorphous materials, however, the absence of long-range translational order removes this geometrical framework and makes the identification of analogous defect descriptors considerably more challenging. This difficulty has motivated the search for alternative approaches to characterizing defects in glasses, a problem that remains at the heart of modern theories of amorphous plasticity.

%\subsection{From Crystals to Glasses}
\subsection{From Crystalline Defects to Amorphous Plasticity}\label{sec1_sub_from_crystals_to_glasses}
The success of defect theory in crystalline solids relies fundamentally on the existence of an ordered reference structure. Dislocations, disclinations, and their associated topological charges are defined relative to the translational and rotational symmetries of the crystal lattice. In glasses and other amorphous materials, however, this reference framework is absent. Although these systems behave mechanically as solids and exhibit elasticity, yielding, plastic flow, and fracture, their constituent particles lack long-range translational order. Consequently, the identification of microscopic defects analogous to those found in crystals has remained a longstanding challenge in the physics of amorphous matter \cite{anderson1995,angell1995,debenedetti2001,berthier2011}.

The absence of a well-defined lattice does not imply the absence of structure. On the contrary, numerous studies have demonstrated that amorphous materials possess complex forms of short-range and medium-range organization that strongly influence their dynamical and mechanical properties \cite{ediger2000,debenedetti2001,berthier2011}. Nevertheless, establishing a direct connection between such structural features and macroscopic behavior has proven considerably more difficult than in crystalline systems. Unlike a dislocation, whose geometry and topological character can be identified unambiguously, structural heterogeneities in glasses are often subtle, spatially distributed, and strongly dependent on the observable under consideration.

This difficulty has motivated decades of research aimed at identifying the microscopic entities responsible for relaxation and plastic deformation in amorphous solids. Various approaches have been proposed, ranging from local free-volume fluctuations and localized shear transformations to geometrical descriptions based on locally favored structures and bond-orientational order \cite{argon1979,falk1998,tanaka2010,tanaka2019}. Experimental observations of localized rearrangements in colloidal glasses have further reinforced the view that plasticity originates from spatially heterogeneous regions embedded within an otherwise elastic matrix \cite{schall2007}. Despite substantial progress, no single structural descriptor has achieved the same level of universality and predictive power as the defect concepts that underpin the mechanics of crystalline materials.

The search for defects in glasses has therefore followed a different trajectory from that in crystalline solids. Rather than identifying singularities in an underlying lattice, researchers have increasingly focused on localized rearrangements, structural heterogeneities, soft regions, and collective dynamical processes. These efforts have led to a variety of complementary theoretical frameworks that seek to describe the microscopic origin of plasticity and mechanical failure in amorphous materials. We briefly review the most influential of these approaches in the following subsection before turning to more recent developments based on non-affine deformation fields, vibrational modes, and topological characterizations.

\subsection{Classical Theories of Plasticity in Glasses}\label{sec1_sub3}
A variety of theoretical frameworks aimed at identifying the microscopic origins of plastic deformation in amorphous solids. Although these approaches differ significantly in their underlying assumptions and methodologies, they share a common objective: to determine whether specific structural or dynamical entities play a role analogous to that of dislocations in crystals. Over the past several decades, several influential ideas have emerged, including free-volume theories, shear transformation zones, geometric frustration and locally favored structures, and descriptions based on elastic inclusions and long-range stress redistribution. Together, these approaches have shaped the modern understanding of plasticity in glasses and continue to provide important conceptual foundations for contemporary research.

\subsubsection{Free-volume theory}\label{sec1_sub3_sub1}
One of the earliest attempts to explain the dynamics and deformation of amorphous materials was based on the concept of free volume. In their pioneering work, Cohen and Turnbull proposed that structural relaxation in dense disordered systems is controlled by the local availability of excess volume, which enables particles to escape their cages and undergo rearrangements \cite{cohen1959,turnbull1961}. Within this picture, molecular mobility is governed not only by thermal activation but also by the distribution of free volume throughout the material. Regions containing larger free volume are expected to rearrange more readily, leading to spatially heterogeneous dynamics and flow.

The free-volume concept was later extended to plastic deformation in metallic glasses by Spaepen, who argued that irreversible shear deformation proceeds through local fluctuations in free volume that facilitate atomic rearrangements under stress \cite{spaepen1977}. These ideas provided one of the first microscopic frameworks for understanding plastic flow in amorphous materials and established the notion that localized structural heterogeneities play a crucial role in deformation. Although free volume remains a useful phenomenological concept, its precise structural definition and direct experimental quantification continue to be subjects of debate.

\subsubsection{Shear transformation zones}\label{sec1_sub3_sub2}

A major advance in the understanding of amorphous plasticity came from the recognition that deformation is highly localized at the microscopic scale. Building on observations in metallic glasses, Argon proposed that plastic flow originates from localized clusters of atoms undergoing cooperative shear-like rearrangements \cite{argon1979}. These elementary events were later formalized within the shear transformation zone (STZ) theory developed by Falk and Langer \cite{falk1998}. In this framework, plastic deformation is mediated by localized regions that can switch between distinct configurational states under applied stress, thereby producing irreversible strain.

STZ theory provided a physically transparent description of yielding and flow in amorphous solids and established a direct connection between microscopic rearrangements and macroscopic constitutive behavior \cite{falk1998,langer2008}. The framework has subsequently been extended to account for a wide range of phenomena, including strain localization, shear-band formation, and rheological response under nonequilibrium conditions \cite{manning2007}. Experimental studies of metallic glasses and colloidal glasses have provided strong evidence for the existence of localized rearrangements consistent with the STZ picture \cite{greer1995,schuh2007,wang2012,johnson2005,wang2015,schall2007}. Nevertheless, an important open question concerns the structural origin of STZs themselves. While the theory successfully describes the dynamics of localized plastic events, identifying the structural features that determine where such events occur remains an active area of investigation.

\subsubsection{Geometric frustration and locally favored structures}\label{sec1_sub3_sub3}
A complementary line of research seeks to identify structural motifs that are intrinsically favored at the local scale but cannot tile Euclidean space without introducing frustration. The origins of this idea can be traced to Frank's seminal observation that icosahedral arrangements possess a lower local energy than competing crystalline structures in simple liquids \cite{frank1952}. Because perfect icosahedral order is incompatible with translational periodicity in three-dimensional Euclidean space, its proliferation necessarily generates geometrical frustration. This observation motivated the proposal that frustration may play a central role in the formation of amorphous solids and the suppression of crystallization.

These ideas were subsequently developed within a broader topological framework by Nelson, Steinhardt, and co-workers, who emphasized the role of disclination defects and frustrated local order in determining the structure and dynamics of supercooled liquids and glasses \cite{nelson1983,nelson1985,steinhardt1983}. In this picture, the inability of locally preferred structures to propagate throughout the system leads to a competition between local ordering tendencies and global geometric constraints. Related theoretical developments by Kivelson, Tarjus, and collaborators established geometric frustration as a possible organizing principle for understanding slow dynamics and the glass transition \cite{kivelson1995,tarjus1995,tarjus2005}.

Subsequent experimental and numerical studies have provided substantial evidence for the presence of locally favored structures in a variety of glass-forming systems. In particular, icosahedral and related motifs have been identified in metallic glasses and supercooled liquids, where they often exhibit strong correlations with dynamical slowing down and enhanced local stability \cite{miracle2004,sheng2006,hirata2013,coslovich2011,Royall13,royall2015}. More recently, bond-orientational order and medium-range structural correlations have been shown to play an important role in describing structural heterogeneity in glass-forming liquids \cite{tanaka2010,tanaka2019,royall2015}. Despite these advances, the relationship between locally favored structures and plastic deformation remains subtle. While such motifs are often associated with increased mechanical stability, they do not provide a universally accepted defect description analogous to the dislocation framework of crystalline solids.

\subsubsection{Eshelby inclusions and elastic interactions}\label{sec1_sub3_sub4}
A different perspective on amorphous plasticity emerged from continuum elasticity. In his pioneering work, Eshelby derived the elastic field generated by a localized transformation embedded within an elastic medium, demonstrating that a local inclusion induces long-ranged stress and displacement fields throughout the surrounding material \cite{eshelby1957,eshelby1959,mura1987}. Although originally developed within the context of continuum elasticity, Eshelby's solution later proved highly relevant for understanding localized plastic rearrangements in amorphous solids.

\begin{figure*}[t]
\centering
\includegraphics[width=\textwidth]{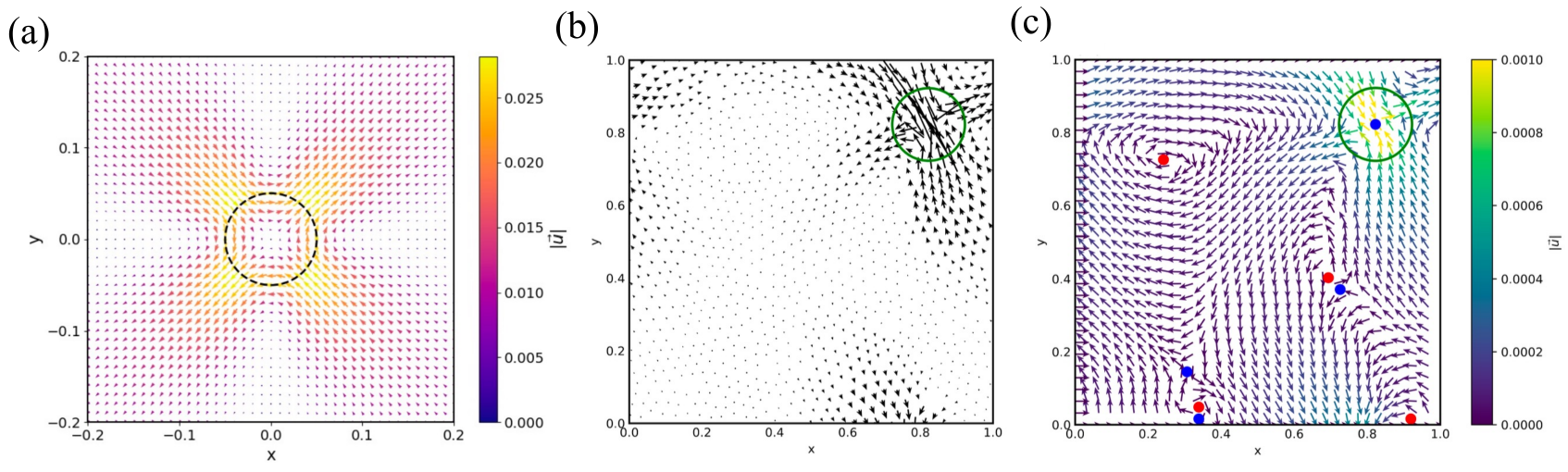}
\caption{Topological structure of Eshelby-like plastic rearrangements.
(\textbf{a}) Analytical displacement field generated by an isolated Eshelby inclusion centered at the origin. The dashed circle denotes the inclusion boundary, while the color scale represents the displacement magnitude. (\textbf{b}) Displacement field obtained from a representative particle-based simulation. The highlighted region identifies a localized Eshelby-like rearrangement embedded within the surrounding elastic medium. (\textbf{c}) Corresponding normalized displacement field on a coarse-grained grid. Positive and negative topological defects, identified through the winding number of the displacement orientation, are shown in red and blue, respectively. The characteristic arrangement of defects reveals the underlying topological structure associated with the Eshelby-like event. Adopted from Bera \emph{et al.}, arXiv:2505.23069 (2025) \cite{bera2025burgersrings}.}
\label{sec1_fig_eshelby}
\end{figure*}

Atomistic simulations revealed that elementary plastic events in glasses are accompanied by displacement fields exhibiting a quadrupolar symmetry closely resembling the elastic response predicted by Eshelby's theory \cite{maloney2004,maloney2006,tanguy2006}. 
More recently, particle-scale analyses of jammed solids have provided new insights into the microscopic origin of these quadrupolar displacement fields, demonstrating how local force imbalances and contact-network geometry give rise to the characteristic Eshelby-like non-affine response observed during deformation \cite{willmarth2025}.
These observations led to the view that localized shear transformations can be regarded as effective Eshelby inclusions embedded within an elastic matrix. Building upon this idea, mesoscale elastoplastic models were developed in which plastic activity is represented by localized yielding events that interact through long-ranged elastic stress redistribution \cite{picard2004,bouchbinder2007}. Such models successfully reproduce a wide range of phenomena, including strain localization, avalanche dynamics, and shear-band formation. In Fig.~\ref{sec1_fig_eshelby}(a), we show the displacement field generated by an ideal Eshelby inclusion, which serves as a fundamental continuum description of a localized plastic rearrangement embedded within an elastic medium. Similar quadrupolar displacement patterns are observed in particle-based simulations of glasses, as illustrated in Fig.~\ref{sec1_fig_eshelby}(b), providing evidence that plastic events in amorphous solids often exhibit an Eshelby-like character \cite{bera2025burgersrings}. The corresponding topological defects identified from the displacement field are shown in Fig.~\ref{sec1_fig_eshelby}(c). Their significance and connection to plasticity will be discussed in detail in later sections of this review.

The Eshelby paradigm has subsequently become one of the central frameworks for understanding amorphous plasticity. Numerical studies have clarified the role of elastic interactions in organizing collective rearrangements and driving the emergence of system-spanning deformation patterns \cite{lerner2009,sengupta2012,dasgupta2013,hentschel2015}. At the same time, the identification of the microscopic locations at which Eshelby-like transformations nucleate remains an important open problem. In contrast to crystalline solids, where defects can be defined relative to an underlying lattice, the structural origin of localized plastic rearrangements in glasses remains less clearly established. This challenge has motivated the search for alternative structural, mechanical, and topological descriptors capable of predicting plastic activity before deformation occurs.

\subsection{Non-affine Motion and Vibrational Modes}\label{sec1_sub5}
While free-volume theories, shear transformation zones, and geometric frustration provide valuable insights into the mechanisms of amorphous plasticity, they do not directly address how structural disorder modifies the elastic response of a solid. In crystalline materials, an externally imposed deformation is transmitted through the lattice in an essentially affine manner, meaning that particle displacements follow the macroscopic strain field. In contrast, the lack of local inversion symmetry in amorphous solids generates additional particle motions that cannot be described by the imposed affine deformation alone. These non-affine displacements play a central role in determining the mechanical and vibrational properties of disordered materials \cite{alexander1998,tanguy2002,leonforte2005,didonna2005,ellenbroek2006,zaccone2011,zaccone2013,zaccone2023book}.

Consider a system subjected to an infinitesimal strain increment. The displacement of particle $i$ can be decomposed into an affine contribution prescribed by continuum elasticity and a non-affine correction,
\begin{equation}
\mathbf{u}_i=\mathbf{u}_i^{\mathrm{A}}+\mathbf{u}_i^{\mathrm{NA}},
\end{equation}
where $\mathbf{u}_i^{\mathrm{A}}$ denotes the affine displacement and $\mathbf{u}_i^{\mathrm{NA}}$ represents the additional relaxation required to restore local mechanical equilibrium. In disordered solids, these non-affine motions can be comparable in magnitude to the affine contribution and lead to substantial corrections to the elastic constants \cite{lemaitre2006,zaccone2011,zaccone2013}. The non-affine theory of elasticity developed by Zaccone and Scossa-Romano established that the elastic moduli can be expressed as the difference between an affine (Born) contribution and a non-affine correction arising from disorder-induced force imbalances \cite{zaccone2011,zaccone2013,zaccone2023book}.

The microscopic origin of non-affinity is naturally described through the Hessian matrix, which characterizes the local curvature of the potential-energy landscape. For a system of $N$ particles interacting through a potential energy $U$, the Hessian is defined as
\begin{equation}
H_{ij}^{\alpha\beta}=\frac{\partial^2 U}{\partial r_i^\alpha \partial r_j^\beta},
\end{equation}
where $r_i^\alpha$ denotes the $\alpha$-th Cartesian coordinate of particle $i$. The vibrational modes of the system are obtained from the eigenvalue problem
\begin{equation}
\mathbf{H}\mathbf{e}_n=\lambda_n \mathbf{e}_n,
\end{equation}
where $\mathbf{e}_n$ is the eigenvector associated with eigenvalue $\lambda_n=\omega_n^2$. The set of eigenvectors forms a complete basis for describing small-amplitude excitations around a mechanically stable configuration.

Studies of the vibrational spectrum of glasses revealed the presence of low-frequency modes that differ qualitatively from ordinary phonons and often exhibit a pronounced degree of spatial localization \cite{wyart2005,wyart2010,lerner2014}. These modes are closely related to regions that are mechanically soft and susceptible to rearrangement under external loading. Early simulations demonstrated strong correlations between dynamic heterogeneity and localized soft regions in supercooled liquids \cite{widmercooper2004,widmercooper2006,widmercooper2008}. Building on these observations, Manning and co-workers introduced the concept of \emph{soft spots}, namely localized regions identified from low-frequency quasi-localized vibrational modes that exhibit an enhanced propensity for future plastic rearrangements \cite{manning2011}. Similar ideas have subsequently been extended using machine-learning approaches that combine structural descriptors with vibrational information to predict local plastic activity \cite{schoenholz2014,cubuk2015,cubuk2016,cubuk2017,richard2020}.

These developments established a direct connection between the vibrational properties of amorphous solids and their mechanical response. Rather than searching for defects directly in the particle configuration, one may characterize the system through fields derived from non-affine displacements or vibrational eigenvectors. As we discuss in the following subsection, such vector fields can exhibit singular structures possessing well-defined topological properties. This observation has opened a new route toward identifying and characterizing defects in glasses, providing a bridge between amorphous plasticity, elasticity, and topology \cite{lerner2016_PRE,Gartner}.

\subsection{Topological Defects in Vector Fields}\label{sec1_sub6}
TDs are singular structures that arise in continuous fields when a globally smooth ordering cannot be maintained throughout the system. Unlike conventional defects defined through local structural irregularities, TDs are characterized by quantities that remain invariant under continuous deformations of the underlying field. Such defects play a central role in a wide variety of physical systems, including liquid crystals, magnetic materials, superfluids, superconductors, active matter, and synchronization phenomena \cite{kleman1977,kleman2008,mermin1979,chaikin1995}. Their importance stems from the fact that large-scale physical behavior is often governed not by microscopic details, but by the topology, interactions, and dynamics of the defects themselves.
\begin{figure*}[t]
\centering
\includegraphics[width=\textwidth]{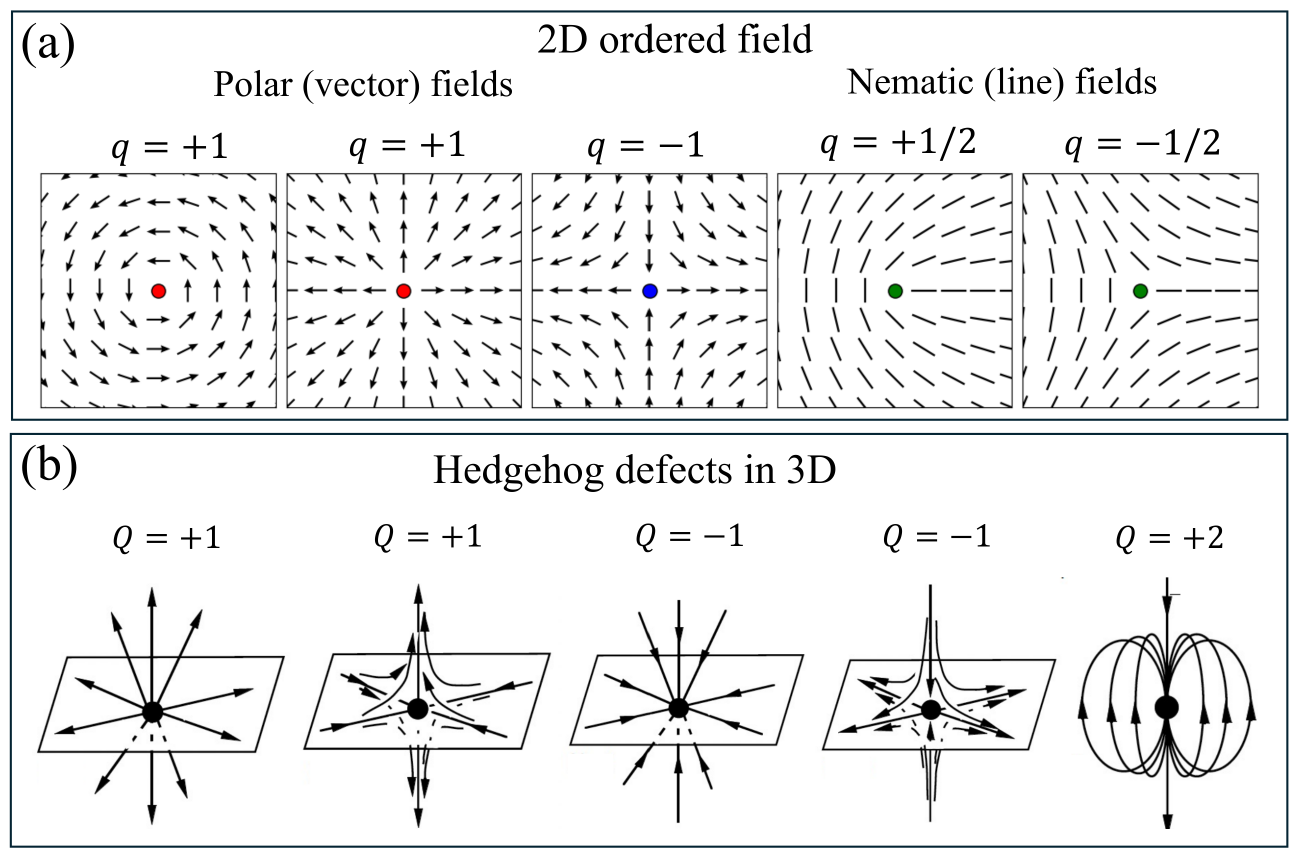}
\caption{Representative topological defects in two and three spatial dimensions.
(\textbf{a}) Examples of point defects in two-dimensional vector and nematic fields. For polar vector fields, a vortex (\(q=+1\)), a source (\(q=+1\)), and an anti-vortex (\(q=-1\)) are shown, where the red (blue) dot indicates the defect core. For nematic fields, the characteristic comet-shaped \(+1/2\) defect and triangular \(-1/2\) defect are illustrated. These defects are characterized by their winding number \(q\), which measures the total rotation of the local orientation field around the defect core.
(\textbf{b}) Schematic examples of three-dimensional point defects (hedgehogs) carrying different topological charges \(Q\). Panel (b) adapted with permission from Kl\'eman and Lavrentovich, in the chapter \emph{Topological Theory of Defects} of the book \emph{Soft Matter Physics: An Introduction} (Springer, New York, 2003), Fig.~12.22~\cite{kleman2003book}. Copyright \textcopyright\ 2003 Springer Nature.}
\label{sec1_fig2}
\end{figure*}

\subsubsection{Two-dimensional vortices and anti-vortices}\label{sec1_sub6_sub1}
The simplest TDs arise in two-dimensional vector fields. Consider a normalized vector field $\mathbf{v}(\mathbf{r})$ characterized by a local orientation angle $\theta(\mathbf{r})$. The topological charge, or winding number, associated with a singular point is defined as \cite{kleman2003book}
\begin{equation}
q=\frac{1}{2\pi}
\oint_{\mathcal{C}} d\theta ,
\label{eq_winding_2d}
\end{equation}
where the contour $\mathcal{C}$ encloses the defect. The winding number measures the total rotation of the vector field around the singularity. A defect with $q=+1$ corresponds to a vortex, whereas a defect with $q=-1$ corresponds to an anti-vortex. More generally, systems possessing additional symmetries may support fractional topological charges. In Fig.~\ref{sec1_fig2}(a), we show two-dimensional vector fields with different TDs characterized by their winding number $q$. These include vortices (source, sink, circulation), anti-vortices in polar vector fields, as well as the familiar $q=+1/2$ and $q=-1/2$ defects observed in nematic systems \cite{shankar2022}.

Vortices and anti-vortices constitute fundamental excitations in numerous physical systems. In the two-dimensional XY model, they govern the celebrated Berezinskii--Kosterlitz--Thouless (BKT) transition through the binding and unbinding of vortex--anti-vortex pairs \cite{kosterlitz1973,kosterlitz1974}. Similar topological excitations appear in superfluids and superconductors, where quantized vortices play a central role in determining the macroscopic response \cite{onsager1949,feynman1955}. Related structures have also been identified in synchronization phenomena described by the Kuramoto model and in active matter systems exhibiting collective motion and self-organization \cite{kuramoto1984,shankar2022}.

An important feature of two-dimensional TDs is that they interact through a long-ranged logarithmic potential \cite{kosterlitz1973,kosterlitz1974},
\begin{equation}
U(q_i,q_j,r_{ij})=-2\pi K\, q_i q_j\ln\!\left(\frac{r_{ij}}{a}\right),
\label{eq:vortex_interaction}
\end{equation}
where $K$ denotes the elastic stiffness, $a$ is a microscopic core cutoff, and $r_{ij}$ is the separation between defects carrying charges $q_i$ and $q_j$. This interaction is formally equivalent to the Coulomb interaction between charges in two-dimensional electrostatics. Defects with opposite charges attract and may annihilate, whereas defects carrying the same charge repel one another. The collective dynamics of these interacting topological charges often dominate the large-scale behavior of the system.

\subsubsection{Three-dimensional hedgehog defects}\label{sec1_sub6_sub2}

The notion of TDs can be generalized to three-dimensional vector fields. In this case, point defects known as hedgehogs arise when vectors radiate outward from or inward toward a singular point. A radial source corresponds to a positive hedgehog, whereas a radial sink corresponds to a negative hedgehog. Such defects are commonly encountered in liquid crystals, magnetic systems, and active matter \cite{kleman2008,mermin1979,chaikin1995}.
A particularly striking manifestation of defect-mediated symmetry breaking was reported early on by Lavrentovich and Terentjev, who demonstrated both experimentally and theoretically a temperature-driven transition between radial and hyperbolic hedgehog configurations in nematic droplets. Remarkably, the transition occurred without any change in the topological charge of the defect and was driven solely by changes in the relative Frank elastic constants, providing one of the earliest examples of a phase transition involving only the symmetry of a topological defect \cite{eugene1986}.

For a unit vector field $\mathbf{s}(\mathbf{r})$, the hedgehog charge is given by \cite{kleman2003book}
\begin{equation}
Q=\frac{1}{4\pi} \int_{\Sigma}d^2x \; \mathbf{s}\cdot \left(\partial_1 \mathbf{s} \times \partial_2 \mathbf{s} \right),
\label{eq_winding_3d}
\end{equation}
where $\Sigma$ denotes a closed surface surrounding the defect. The integer-valued quantity $Q$ measures how the vector field maps the enclosing surface onto the unit sphere and remains invariant under smooth deformations of the field. Unlike the winding number in two dimensions, which characterizes the rotation of a field around a closed loop, the hedgehog charge quantifies the global orientational structure of a three-dimensional vector field surrounding a singular point. Representative examples are shown in Fig.~\ref{sec1_fig2}(b), where point-like hedgehog defects carrying different topological charges $Q$ are illustrated. In contrast to two-dimensional defects, where opposite circulation directions or source--sink configurations carry the same winding number $q$, in three dimensions reversing the direction of the vector field changes the sign of the topological charge $Q$. More generally, three-dimensional systems can support a wider variety of topological excitations, including point defects and line defects, and in certain systems also defect surfaces or interfaces \cite{mermin1979,chaikin1995}. In the present discussion, however, we focus exclusively on point-like singularities. In this context, hedgehog defects provide the natural three-dimensional counterpart of vortices and anti-vortices and play a central role in the topological description of vector field.

\subsubsection{Topological defects in glasses}\label{sec1_sub6_sub3}
Although topological defects have long been studied in ordered systems, their relevance to amorphous solids has only recently begun to attract significant attention. Interestingly, vortex-like displacement patterns have been observed for many years in simulations of non-affine deformation and low-frequency vibrational modes of glasses \cite{tanguy2002,leonforte2005,didonna2005,ellenbroek2006}. These studies revealed highly heterogeneous displacement fields containing rotational structures reminiscent of vortices in conventional vector fields, although their topological nature was not explicitly analyzed.

More recently, these observations have been revisited from a topological perspective. Vortices, anti-vortices, and hedgehog defects have been identified in vector fields derived from non-affine displacements and from the eigenvectors of the Hessian matrix, enabling a systematic characterization of topological charge distributions in amorphous solids \cite{baggioli2021,wu2023,desmarchelier2024}. These developments suggest that topological defects may provide a new framework for describing structural and mechanical heterogeneities in glasses and may offer an alternative route toward identifying the microscopic precursors of plastic activity.

At present, however, the precise relationship between topological defects, localized plastic rearrangements, soft spots, and the underlying structural organization of amorphous materials remains incompletely understood. Whether these defects constitute the amorphous analogue of crystalline defects, or instead represent a complementary description of mechanical heterogeneity, remains an open question. Addressing this issue is one of the central objectives of the present review.

\section{Topological Defects in Simulations of Amorphous Solids}\label{section2}
The theoretical ideas reviewed in the previous section provide the conceptual foundations for understanding topological defects in amorphous materials. A central challenge, however, is how such defects can be identified and characterized in systems that lack an underlying crystalline lattice. Over the last few years, advances in numerical simulations have led to the emergence of a new framework in which topological defects are detected not from the particle configuration itself, but from vector fields constructed from mechanical observables. These include vibrational eigenmodes obtained from the Hessian matrix, non-affine displacement fields generated during deformation, and coarse-grained fields associated with localized plastic rearrangements. Remarkably, these fields exhibit well-defined vortices, anti-vortices, and hedgehog defects whose spatial organization carries important information about mechanical stability, plastic activity, yielding, and shear-band formation.

In this section, we review recent simulation studies that have established topological defects as a useful framework for describing the mechanical response of glasses. We begin with vibrational eigenmodes, where the first systematic identification of topological defects in amorphous solids was achieved. We then discuss the extension of these ideas to non-affine displacement fields, localized plastic rearrangements, collective defect organization, shear-band formation, Burgers-ring structures, and three-dimensional hedgehog defects. Together, these developments reveal a growing body of evidence that topology provides a complementary perspective on the long-standing problem of identifying the elementary carriers of plasticity in amorphous solids.

\subsection{Topology of Vibrational Eigenmodes}\label{sec2_sub1}
The first systematic identification of topological defects in amorphous solids was achieved through the analysis of vibrational eigenmodes of the Hessian matrix \cite{wu2023}. As discussed in Sec.~\ref{section1}, the Hessian matrix characterizes the local curvature of the potential-energy landscape around a mechanically stable configuration. Its eigenvectors provide a complete basis for describing infinitesimal deformations and vibrational excitations of the system. Traditionally, studies of vibrational modes in glasses have focused on quantities such as the density of states, participation ratio, and localization properties. The work of Wu \emph{et al.} \cite{wu2023} approached the problem from a different perspective by treating the eigenvector field itself as a quantity that can possess nontrivial topological structure.

\begin{figure*}[t]
\centering
\includegraphics[width=\textwidth]{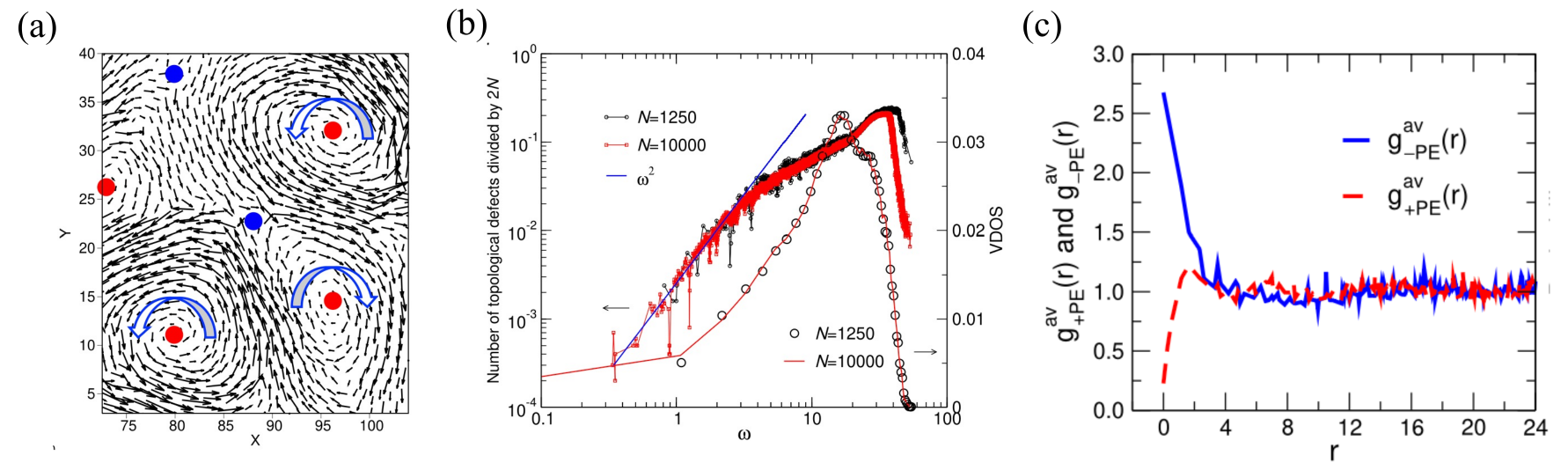}
\caption{Topological defects in vibrational eigenmodes of amorphous solids and their connection to plastic activity. (\textbf{a}) Representative low-frequency eigenvector field ($\omega=1.5$) obtained from the Hessian matrix of a two-dimensional glass, illustrating the emergence of vortex- and anti-vortex-like structures in the vibrational mode.
(\textbf{b}) Frequency dependence of the topological defect density (left axis) together with the vibrational density of states $D(\omega)$ (right axis) for two system sizes, highlighting the relation between topological excitations and low-frequency vibrational modes.
(\textbf{c}) Correlation between the spatial locations of topological defects and subsequent plastic rearrangements, demonstrating the predictive capability of vibrational-mode topology for identifying mechanically susceptible regions. Adapted from Wu \emph{et al.}, Nat. Commun. \textbf{14}, 2955 (2023) \cite{wu2023}, licensed under CC BY 4.0.}
\label{sec2_fig1}
\end{figure*}

For a given vibrational mode, each particle is associated with an eigenvector that specifies the direction and amplitude. By considering the local orientation of these vectors, one may define a continuous vector field and compute the winding number around a closed contour according to Eq. \ref{eq_winding_2d}. Remarkably, the eigenmodes were found to host numerous vortices and anti-vortices carrying topological charges $q=\pm1$, despite the absence of any underlying crystalline order. A zoom-in portion of vibrational field at low-frequency is shown in Fig.~\ref{sec2_fig1}(a), where the singular points of the eigenvector field can be clearly identified through the circulation of the local orientation \cite{wu2023,bera2024pnasnexus}. In Fig.~\ref{sec2_fig1}(b), the vibrational density of states $D(\omega)$ and the density of topological defects are plotted as functions of the eigenfrequency $\omega$. Remarkably, both quantities exhibit a similar low-frequency scaling behavior, with the defect density increasing approximately as $N_d \sim \omega^2$, mirroring the Debye-like scaling of the vibrational density of states in this regime.

A key result of Ref.~\cite{wu2023} was the observation that these topological defects are not randomly distributed. Instead, they exhibit strong correlations with mechanically soft regions and with the spatial locations where plastic rearrangements occur subsequently. In particular, low-frequency quasi-localized modes were found to contain an enhanced density of topological defects, suggesting a direct connection between vibrational heterogeneity and local mechanical instability. Fig.~\ref{sec2_fig1}(c) shows the average spatial correlation between soft spots and topological defects. At short distances, defects carrying negative topological charge display a noticeably stronger correlation with soft spots than their positively charged counterparts.

One important finding was that topological defects identified in the undeformed glass possess predictive power for future plastic activity. Regions surrounding topological defects display an elevated probability of undergoing irreversible rearrangements under subsequent loading, thereby establishing a direct connection between the topology of vibrational modes and the onset of plastic deformation. These results provided one of the first demonstrations that topological concepts can be used not merely to characterize the structure of a glass, but also to identify mechanically vulnerable regions before plastic events occur.

The work \cite{wu2023} represents an important conceptual advance because it shifted the focus from particle configurations to vector fields derived from mechanical observables. Rather than searching for structural defects in the disordered arrangement of particles, the relevant topological information emerges naturally in the eigenvector fields associated with the vibrational spectrum. This discovery opened a new direction in the study of amorphous solids and laid the foundation for subsequent investigations of topological defects in non-affine displacement fields, localized plastic rearrangements, and yielding phenomena.

\subsection{Topological Defects in Non-affine Displacement Fields}\label{sec2_sub2}
A central feature distinguishing amorphous solids from crystals is the presence of non-affine displacements. In crystalline materials, particle motions under deformation largely follow the imposed macroscopic strain. In contrast, in glasses structural disorder breaks local inversion symmetry and generates additional relaxations that are required to restore mechanical equilibrium \cite{lemaitre2006,zaccone2011,zaccone2013,zaccone2023book}. The total displacement field can therefore be decomposed as
\begin{equation}
u_i(\mathbf{x})=
F_{ij}x_j+ u_i^{\mathrm{NA}}(\mathbf{x}),
\end{equation}
where the first term represents the affine component of the deformation, with $F_{ij}$ denoting the deformation gradient tensor, while $u_i^{\mathrm{NA}}(\mathbf{x})$ describes the non-affine displacement field arising from local structural disorder and particle rearrangements. Here we used Latin indices $ij$ to denote Cartesian components. The non-affine displacement field is highly heterogeneous and encodes the mechanical response of the underlying disordered structure.
\begin{figure*}[t]
\centering
\includegraphics[width=\textwidth]{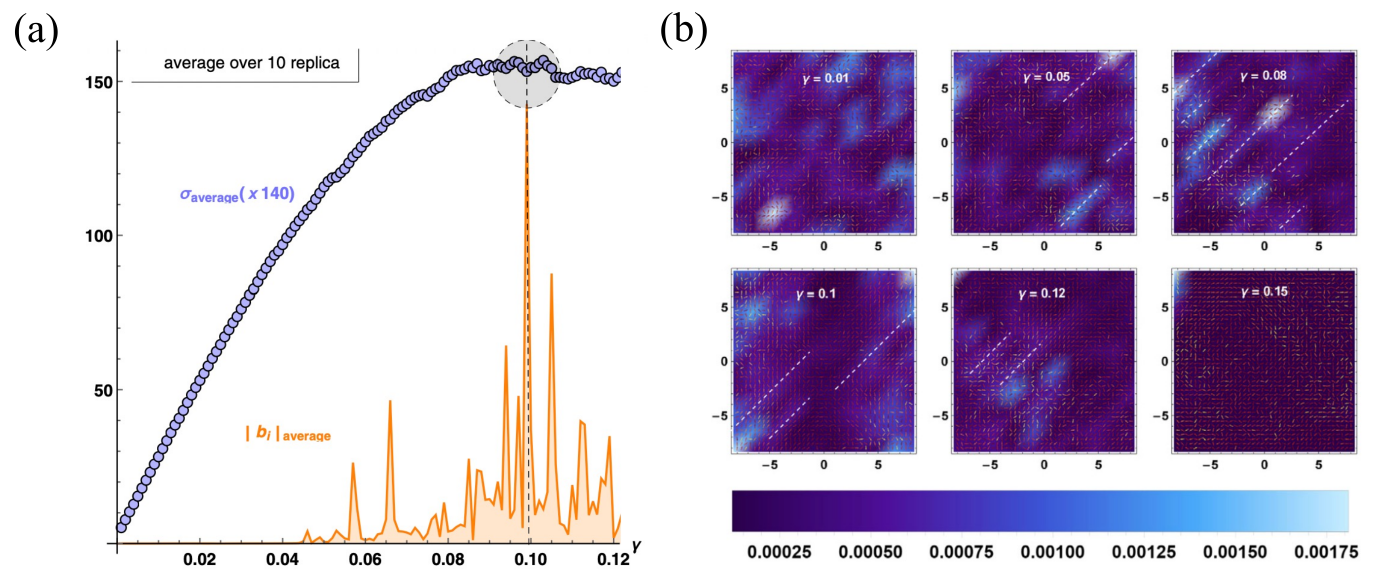}
\caption{Evolution of the generalized Burgers vector during deformation and its connection to strain localization. (\textbf{a}) Stress - strain response of a two-dimensional amorphous solid (purple circles) together with the average magnitude of the Burgers vector, $|\mathbf{b}|$. The pronounced increase in $|\mathbf{b}|$ near the stress-drop event indicates the accumulation of deformation incompatibility preceding plastic relaxation. (\textbf{b}) Spatial evolution of the displacement field under increasing applied strain $\gamma$. Arrows represent the local displacement vectors, while the background color denotes the magnitude of the Burgers field. Regions of enhanced Burgers magnitude progressively organize into a localized deformation pattern aligned approximately along the principal shear direction, ultimately developing into a shear band. Panels adapted with permission from Baggioli \emph{et al.}, Phys. Rev. Lett. \textbf{127}, 015501 (2021) \cite{baggioli2021}. Copyright (2021) American Physical Society.}
\label{sec2_fig2}
\end{figure*}

Building on these ideas, Baggioli \emph{et al.} proposed a topological characterization of plasticity based on the non-affine displacement field \cite{baggioli2021}. The central idea is that deformation incompatibilities in the non-affine field can be quantified through a generalized Burgers vector, defined by Eq.~\ref{eq:burgers}, in direct analogy with the classical Burgers construction used for crystalline dislocations. Although amorphous solids lack a reference lattice, the coarse-grained displacement field remains well defined, allowing this construction to be extended to disordered systems.

Their simulations revealed pronounced spatial heterogeneity in the generalized Burgers field, with large amplitudes localized in regions of intense non-affine activity. As shown in Fig.~\ref{sec2_fig2}(a), the average magnitude of the Burgers vector exhibits intermittent peaks as a function of the applied strain $\gamma$, closely correlating with major plastic rearrangements. Figure~\ref{sec2_fig2}(b) illustrates the evolution of the non-affine displacement field during deformation. With increasing strain, the displacement patterns progressively organize along directions approximately oriented at $45^\circ$ to the applied strain direction, consistent with the emergence of shear-band-like structures.

This work \cite{baggioli2021} represented an explicit attempt to connect Burgers vector formulation applicable for crystalline systems to the plastic deformation in amorphous solids. Unlike traditional structural indicators, the Burgers vector is constructed directly from the mechanical response of the system and therefore provides a bridge between continuum elasticity, non-affine deformation, and localized plastic rearrangements. These results established that topological concepts can be applied not only to vibrational eigenmodes but also to deformation-induced displacement fields, opening the way for subsequent studies of localized plastic events, yielding, and shear-band formation. 

\subsection{Topological Characterization of Localized Plastic Rearrangements}\label{sec2_sub3}

While topological defects in vibrational eigenmodes and non-affine displacement fields have been shown to identify regions prone to plastic rearrangements, an equally important question is how these defects are connected to the plastic events themselves. In amorphous solids, plastic deformation proceeds through highly localized rearrangements involving a small group of particles embedded within an elastically deforming matrix. These shear transformations generate long-ranged displacement fields that closely resemble the Eshelby solution for a localized inclusion in an elastic medium \cite{eshelby1957,eshelby1959,maloney2004,maloney2006,tanguy2006}.

An important advance in this direction was reported by Desmarchelier \emph{et al.} \cite{desmarchelier2024}, who established a direct connection between topological defects in displacement fields and localized shear transformations. Building on the earlier identification of topological defects in amorphous solids, they showed that negatively charged $q=-1$ defects provide a reliable geometrical marker for locating the centers of plastic rearrangements. As illustrated in Fig.~\ref{sec1_fig_eshelby}, the displacement field associated with an Eshelby-like event contains a negative charged $q=-1$ defect positioned at the center of the localized rearrangement.

Beyond locating plastic events, the authors demonstrated that physically meaningful characteristics of the underlying shear transformation can be extracted directly from the displacement field surrounding the defect. In particular, the orientation and magnitude of the effective eigenstrain can be determined either from the local non-affine displacements or through fits to the Eshelby solution. These quantities were shown to provide a reasonable estimate of the stress relaxation measured in molecular-dynamics simulations, thereby establishing a quantitative link between topological defects and the mechanical response of the material.

These findings provide an important bridge between the classical Eshelby description of plasticity and recent topological approaches. Rather than replacing continuum elasticity, the topological framework complements it by providing a robust method for identifying and characterizing localized shear transformations directly from the displacement field. More broadly, this work suggests that topological defects provide a useful framework for identifying and characterizing the elementary plastic rearrangements that govern deformation in amorphous solids.

\begin{figure*}[t]
\centering
\includegraphics[width=\textwidth]{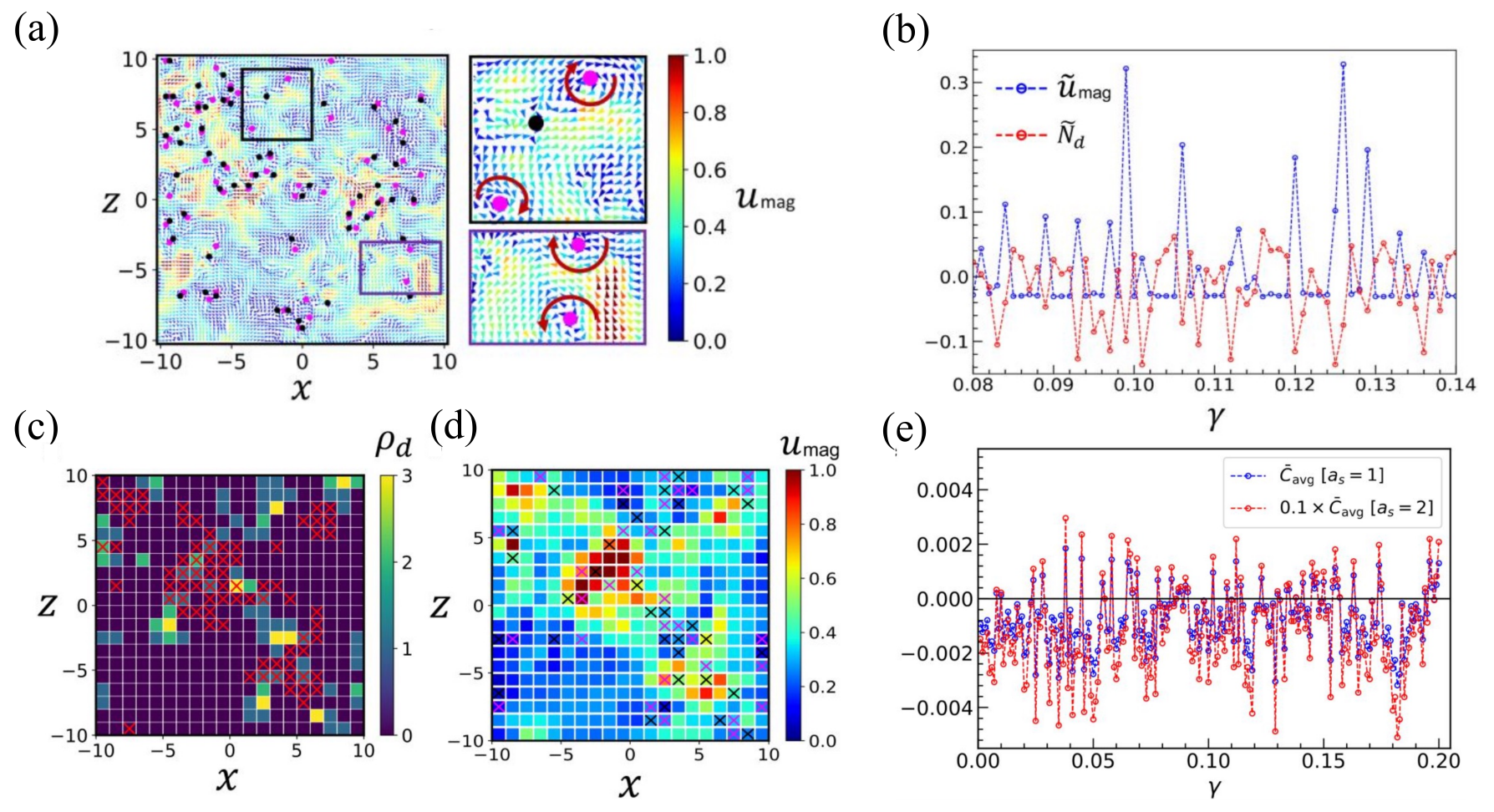}
\caption{Collective organization of topological defects during deformation of a three-dimensional glass.
(\textbf{a}) Non-affine displacement field in an $xz$ slice $y=0$,~$\gamma=0.098$ showing vortices and anti-vortices with topological charges $q=+1$ (magenta) and $q=-1$ (black), respectively. Insets highlight the local chirality of the defects. (\textbf{b}) Evolution of the average non-affine displacement magnitude and the total number of topological defects as a function of strain. (\textbf{c}) Spatial distribution of the local topological charge density for a representative slice near yielding ($\gamma=0.099$); cells exhibiting enhanced non-affine activity are marked by red crosses. (\textbf{d}) The local magnitude of nonaffine displacement is shown in color bar with the cells marked by crosses colored in magenta and black are populated by topological
charges with positive and negative values, respectively. (\textbf{e}) Average topological charge density as a function of strain for two different coarse-graining lengths, illustrating the emergence of correlated defect structures during deformation. Adapted from Bera \emph{et al.}, PNAS Nexus \textbf{3}, pgae315 (2024) \cite{bera2024pnasnexus}, licensed under CC BY.}
\label{sec2_fig3}
\end{figure*}

\subsection{Collective Organization of Topological Defects and Plastic Failure}\label{sec2_sub4}

The studies discussed above establish that topological defects are closely associated with mechanically vulnerable regions and localized plastic rearrangements in amorphous solids. A natural next question is whether the collective organization of these defects carries information about macroscopic failure. In crystalline materials, large-scale plastic deformation emerges from the interactions and accumulation of dislocations. Although amorphous solids lack a crystalline lattice, one may similarly ask whether topological defects in displacement fields exhibit collective behavior as yielding is approached.

This question was investigated by Bera \emph{et al.} \cite{bera2024pnasnexus} using three-dimensional glasses subjected to athermal quasistatic shear. The analysis was performed on two-dimensional slices extracted from the three-dimensional non-affine displacement field. Within each slice, vortices and anti-vortices were identified through the winding number of the coarse-grained displacement field, allowing the evolution of the topological landscape to be monitored throughout deformation. Representative examples are shown in Fig.~\ref{sec2_fig3}(a), where a rich population of positive and negative defects is observed within the displacement field.

A notable finding is that the defect population is strongly coupled to the mechanical response of the material. As shown in Fig.~\ref{sec2_fig3}(b), abrupt increases in the magnitude of the non-affine displacement field, corresponding to plastic events, are accompanied by pronounced variations in the total number of topological defects. This observation indicates that the topological structure of the displacement field undergoes significant reorganization during irreversible rearrangements.

To characterize this behavior quantitatively, Bera \emph{et al.} introduced a coarse-grained topological charge density by partitioning each two-dimensional slice into square cells and computing the net charge within each cell. Representative examples are shown in Fig.~\ref{sec2_fig3}(c,d), where the spatial distribution of local charge density is compared with regions of enhanced non-affine displacement. Cells containing large non-affine activity were found to exhibit a pronounced excess of negative topological charge. This trend is further quantified in Fig.~\ref{sec2_fig3}(e), which reveals a persistent negative charge bias throughout deformation and an increasingly strong association between negatively charged defects and mechanically active regions.

An important consequence of this asymmetry is the emergence of correlated defect structures as the system approaches yielding. While vortices and anti-vortices are present throughout the material, the coarse-grained charge density develops increasingly heterogeneous spatial patterns near major stress-drop events. These observations suggest that macroscopic failure is preceded not only by enhanced local plastic activity but also by a progressive reorganization of the underlying topological landscape.

The results of Ref.~\cite{bera2024pnasnexus} therefore extend the role of topological defects beyond that of local indicators of plastic rearrangements. They demonstrate that the collective distribution of defects contains valuable information about the approach to yielding, with negatively charged defects becoming increasingly concentrated within regions that ultimately participate in plastic failure. This collective organization naturally motivates the investigation of how defect structures evolve into extended patterns during strain localization and shear-band formation, which we discuss in the next subsection.

\subsection{Topological Defects and Shear-Band Formation}\label{sec2_sub5}

Following the onset of yielding, plastic deformation in amorphous solids often becomes localized into narrow regions known as shear bands. These highly concentrated deformation pathways are responsible for catastrophic failure in many glasses and have therefore been the subject of extensive theoretical, numerical, and experimental investigations \cite{argon1979,falk1998,schuh2007,wang2012}. Despite decades of research, the microscopic mechanisms governing the nucleation and growth of shear bands remain incompletely understood.

Recent work by Bera \emph{et al.} \cite{bera2025shearbands,rosner2024} suggests that topological defects provide a useful framework for understanding this phenomenon. Using athermal quasistatic simulations of two-dimensional amorphous solids, the authors analyzed the topology of the non-affine displacement field during large stress-drop events associated with strain localization. Vortices and anti-vortices were identified through the winding number of the displacement field, enabling the spatial organization of topological defects to be tracked throughout the deformation process.

A key observation is that the displacement field contains both trivial and non-trivial topological structures during deformation. Even within the elastic regime, \emph{i.e.}, at small applied strains, a chain of alternating vortices and anti-vortices is present along the center of the simulation box. This background structure arises directly from the imposed shear geometry, since particles above and below the mid-plane move in opposite directions, and therefore represents a trivial topological arrangement. During major plastic events, however, additional non-trivial defect chains emerge that are not constrained by the shear geometry. These structures can appear at arbitrary locations and are typically oriented either horizontally or vertically, depending on the underlying deformation pathway.

\begin{figure*}[t]
\centering
\includegraphics[width=\textwidth]{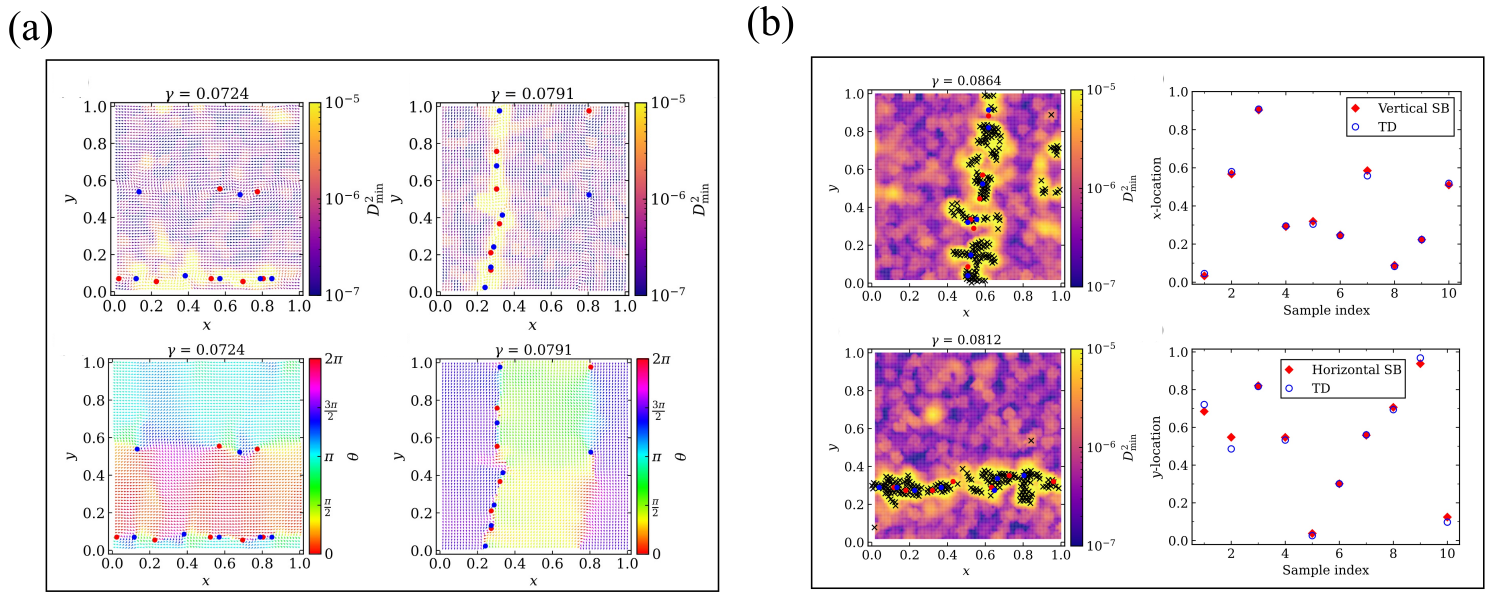}
\caption{Topological organization of defects during shear-band formation in a two-dimensional amorphous solid.
(\textbf{a}) Non-affine displacement fields between consecutive strain increments $\delta\gamma=10^{-4}$, showing vortices ($q=+1$, red) and anti-vortices ($q=-1$, blue) superimposed on the local non-affinity $D_{\mathrm{min}}^2$. Two representative system-spanning shear bands are shown, illustrating the emergence of extended chains of topological defects aligned parallel to the deformation pathway.
(\textbf{b}) Corresponding phase-angle maps of the displacement field, highlighting the defect structures associated with the shear bands and the sharp orientational variations across the defect chains.
(\textbf{c}) Representative vertical shear band at $\gamma=0.0864$ (horizontal shear band at $\gamma=0.0812$). The background color denotes $D_{\mathrm{min}}^2$, while highly non-affine particles and the defects belonging to the principal defect chain are indicated.
(\textbf{d}) Comparison between the average position of the shear band, determined from particles with large $D_{\mathrm{min}}^2$, and the position of the corresponding topological defect chain averaged over multiple independent samples. The close agreement demonstrates that defect chains provide a robust topological signature of shear-band localization. Panel (a) is adapted from Bera \emph{et al.}, arXiv:2507.09250 (2025) \cite{bera2025shearbands}, while panel (b) presents previously unpublished results.}
\label{sec2_fig5}
\end{figure*}

Representative examples are shown in Fig.~\ref{sec2_fig5}(a), where the $D^2_{\rm min}$ field \cite{falk1998} reveals system-spanning shear bands as bright yellow strips. The non-trivial defect chains closely coincide with these regions of intense plastic activity. Owing to the periodic boundary conditions, each primary chain is accompanied by a parallel image chain displaced by approximately $L/2$, although the latter is not associated with significant non-affine activity. The corresponding phase-angle maps shown in the lower panels of Fig.~\ref{sec2_fig5}(a) further highlight the systematic variation of displacement orientation along the defect chains.

The emergence of these non-trivial chains is intimately linked to shear-band formation. Regions exhibiting large non-affine displacements coincide spatially with the primary defect chains, and the average position of these chains accurately tracks the location of the developing shear band. This correspondence is quantified in Fig.~\ref{sec2_fig5}(b), where the position of the shear band, determined independently from the $D^2_{\rm min}$ field, is compared with that of the primary defect chain. The excellent agreement between these two measures indicates that the defect chain typically forms along the center of the shear band. This behavior is consistently observed across independent realizations and for both horizontally and vertically oriented bands, demonstrating its robustness.

From a topological perspective, these results suggest that strain localization is accompanied by the emergence of correlated defect structures beyond the trivial background topology imposed by shear. The shear band is therefore not merely a region of enhanced plastic activity but is also associated with a characteristic arrangement of alternating vortices and anti-vortices embedded within the displacement field. In this picture, defect chains provide a geometrical signature of the pathway along which irreversible rearrangements accumulate and propagate.

Although the precise dynamical mechanisms responsible for the formation of these structures remain to be fully understood, the results of Ref.~\cite{bera2025shearbands} demonstrate that topological defects capture important aspects of the collective organization underlying strain localization. This suggest that the onset of macroscopic failure is accompanied by the emergence of extended topological structures that reflect the spatial organization of plastic activity. Similar observations have also been reported in metallic glasses, where localized vortex-like structure and shear transformation zones (STZ) have been shown to organize collectively during shear-band nucleation and propagation \cite{moshe2015,yu2025,sopu2017,hassani2019}. These studies further support the idea that plastic localization is governed by the spatial organization and interaction of elementary defect-like excitations.

\begin{figure*}[t]
\centering
\includegraphics[width=\textwidth]{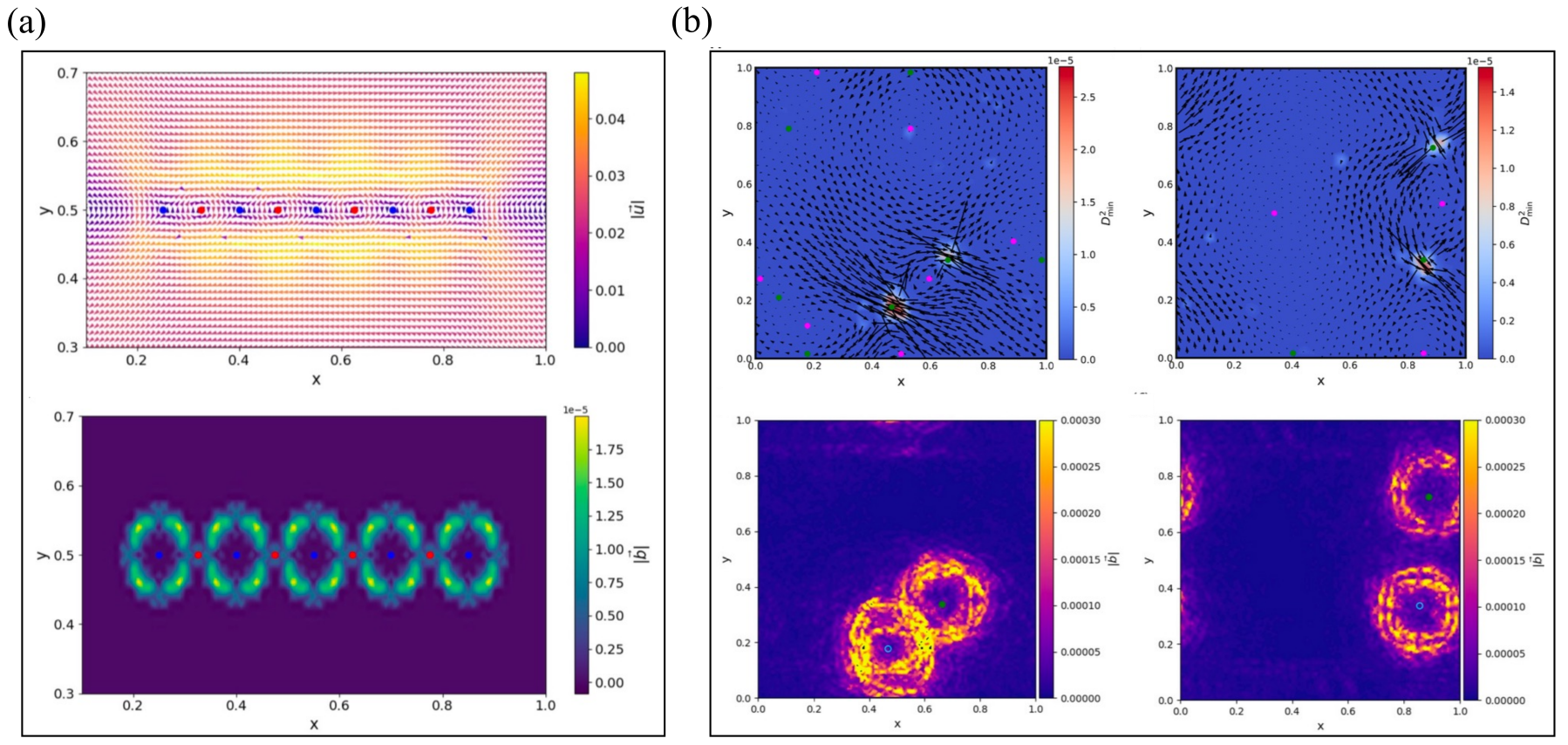}
\caption{Burgers rings as topological signatures of Eshelby-like plastic events in amorphous solids.
(\textbf{a}) (Top) Analytical displacement field generated by five Eshelby inclusions aligned along the \(x\)-direction. Positive and negative winding-number defects are indicated by red and blue symbols, respectively. (Bottom) Corresponding spatial distribution of the continuous Burgers vector magnitude, revealing the emergence of five distinct Burgers rings centered on the Eshelby inclusions.
(\textbf{b}) (Top) Non-affine displacement field containing two localized Eshelby-like rearrangements. The background color represents the local non-affinity \(D_{\mathrm{min}}^2\), highlighting the locations of the underlying plastic events, while positive and negative topological defects are marked by magenta and green circles, respectively. (Bottom) Continuous Burgers vector magnitude for the same configuration, showing the superposition of two Burgers rings associated with the two plastic rearrangements. The anti-vortex defects located at the centers of the rings are highlighted. Adapted from Bera \emph{et al.}, arXiv:2505.23069 (2025) \cite{bera2025burgersrings}.}
\label{sec2_fig6}
\end{figure*}

\begin{figure*}[t]
\centering
\includegraphics[width=\textwidth]{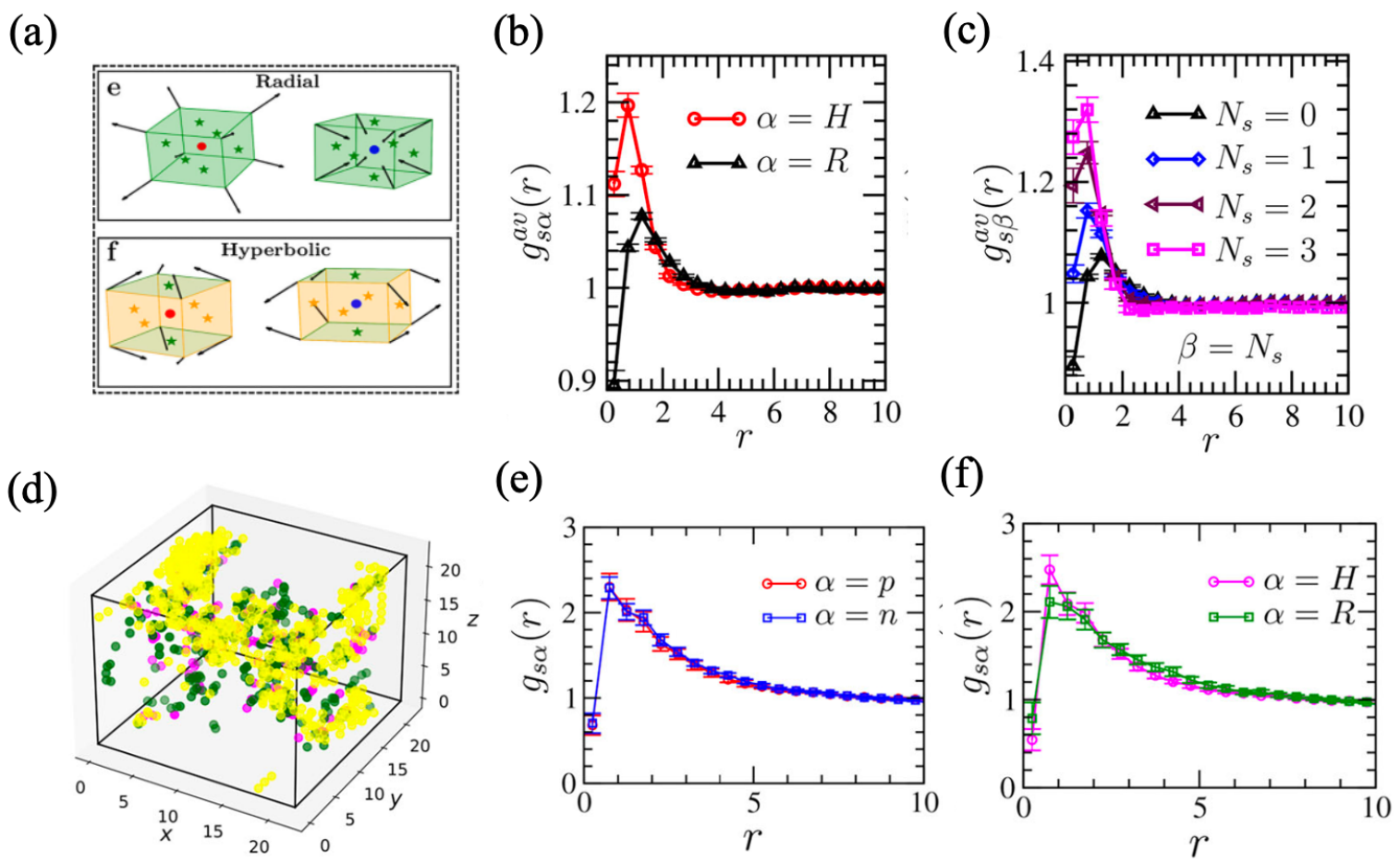}
\caption{Hedgehog topological defects (HTDs) in three-dimensional amorphous solids and their correlation with mechanically soft regions.
(\textbf{a}) Schematic representations of hyperbolic defects with topological charges $Q=\pm 1$, together with the corresponding projected vector fields and winding-number structure.
(\textbf{b}) Average spatial correlation between soft spots and radial (R) and hyperbolic (H) defects, demonstrating a stronger association of soft spots with hyperbolic defects.
(\textbf{c}) Dependence of the soft-spot correlation on the parameter $N_s$, which characterizes the degree of hyperbolic character of the defect.
(\textbf{d}) Representative three-dimensional configuration showing the spatial distribution of hedgehog defects with hyperbolic (magenta) and radial (green) nature and soft spots are indicated by yellow regions. 
(\textbf{e}) Average spatial correlation between soft spots and monopoles $Q=+1$ and anti-monopoles $Q=-1$.
(\textbf{f}) Comparison of the average correlations of hyperbolic and radial defects with soft spots. The statistics are averaged over multiple strain values corresponding to stress-drop events.
Panels adapted from Bera \emph{et al.}, Nat. Commun. \textbf{16}, 5990 (2025) \cite{bera2025hedgehog}.}
\label{sec2_fig4}
\end{figure*}

\begin{figure*}[t]
\centering
\includegraphics[width=\textwidth]{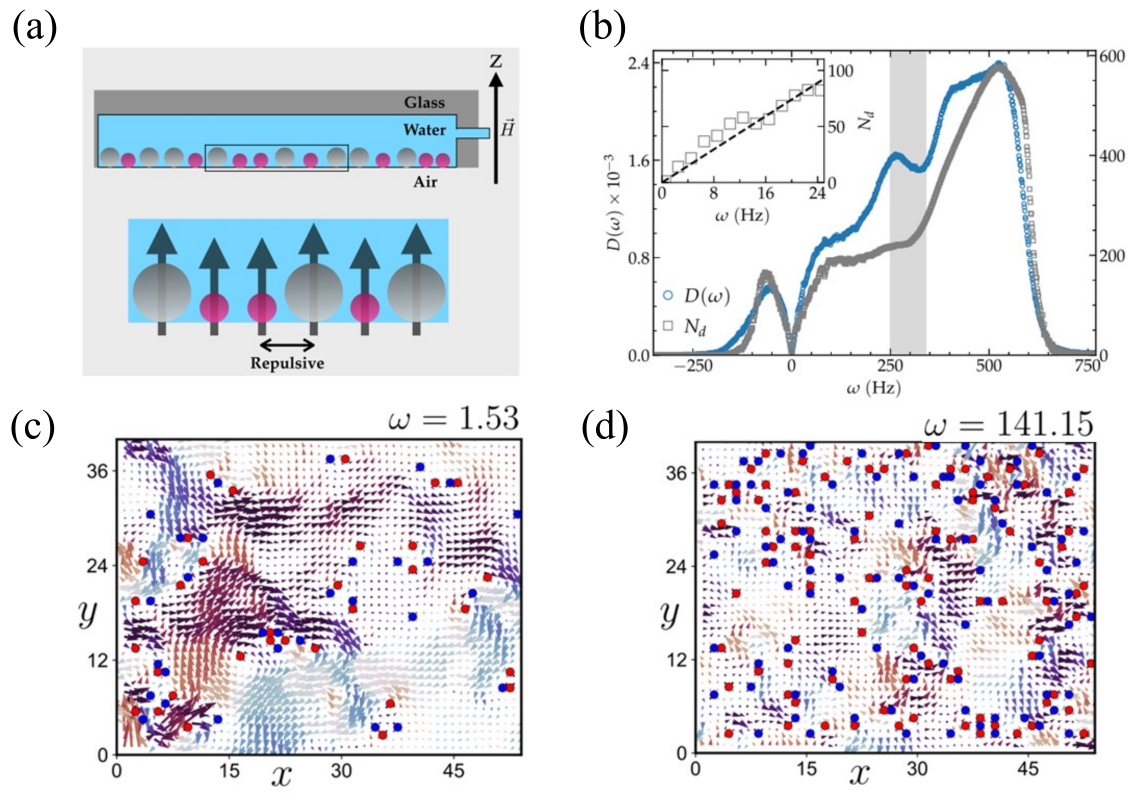}
\caption{Topology of vibrational eigenmodes in a 2D Colloidal Glass. \textbf{(a)} Schematic of the experimental setup: a bidisperse monolayer of superparamagnetic colloidal particles confined at a water--air interface and interacting via tunable dipole--dipole repulsion under an external magnetic field. \textbf{(b)} Vibrational density of states $D(\omega)$ (blue) and total number of topological defects $N_d$ (gray) as a function of frequency, showing a strong correlation between vibrational excitations and defect density. The inset highlights the linear scaling at low frequency. \textbf{(c-d)} Eigenvector fields at low ($\omega=1.53$) and high ($\omega=141.15$) frequencies. Arrows indicate particle displacements, while red and blue dots denote vortices ($q=+1$) and anti-vortices ($q=-1$). Low-frequency modes exhibit coherent, extended swirling patterns with few defects, whereas high-frequency modes are increasingly disordered with a dense defect population. Panels adapted from Vaibhav \emph{et al.} Nat. Commun. \textbf{16}, 55 (2025)~\cite{vaibhav2025}.}
\label{fig:colloidal_topodefects}
\end{figure*}
\begin{figure*}[t]
\centering
\includegraphics[width=\textwidth]{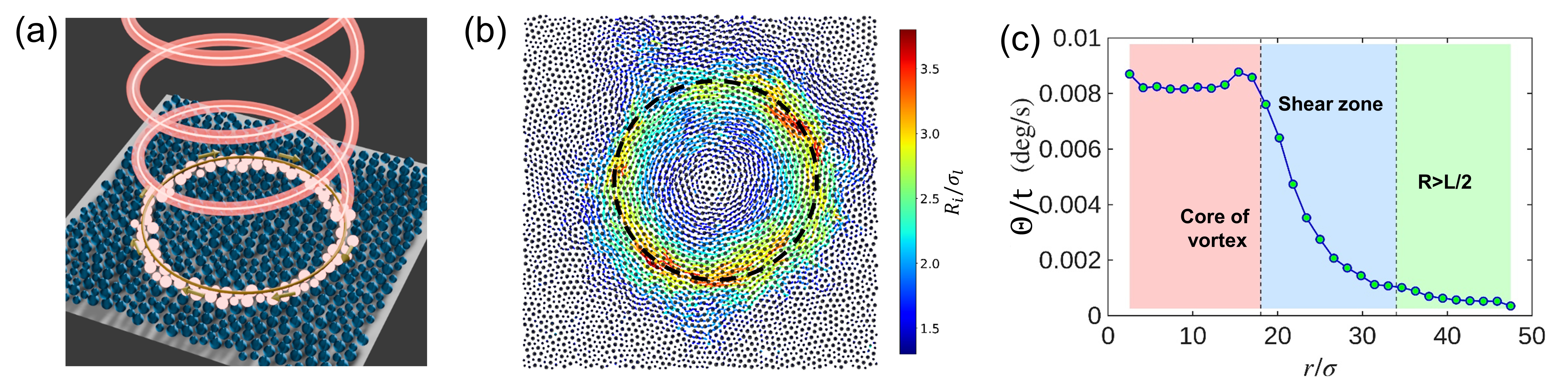}
\caption{A dense colloidal monolayer under shear due to an optical vortex.
(\textbf{a}) Schematic of the experimental setup in which an optical vortex, created using a spatial light modulator (SLM), drives a colloidal monolayer. The blue spheres denote particles of the binary colloidal suspension, whereas the yellow spheres correspond to particles trapped within the optical vortex that acquire orbital angular momentum from the laser beam.
(\textbf{b}) Integrated displacement of particles due to applied shear. Particle trajectories are superimposed on the reconstructed images of the colloidal suspension and are color-coded from blue to red according to the magnitude of displacement. The black circle marks the optical vortex region with a diameter of $30\sigma$.
(\textbf{c}) Different regions of the deformation field are highlighted. The red region denotes the vortex core, corresponding to the area enclosed by the black circle in panel (b). The surrounding blue region represents the shear zone, while the yellow region corresponds to particles located at distances greater than $L/2$ from the vortex center, where $L$ is the size of the field of view.}
\label{S32Fig1}
\end{figure*}

\begin{figure*}[t]
\centering
\includegraphics[width=\textwidth]{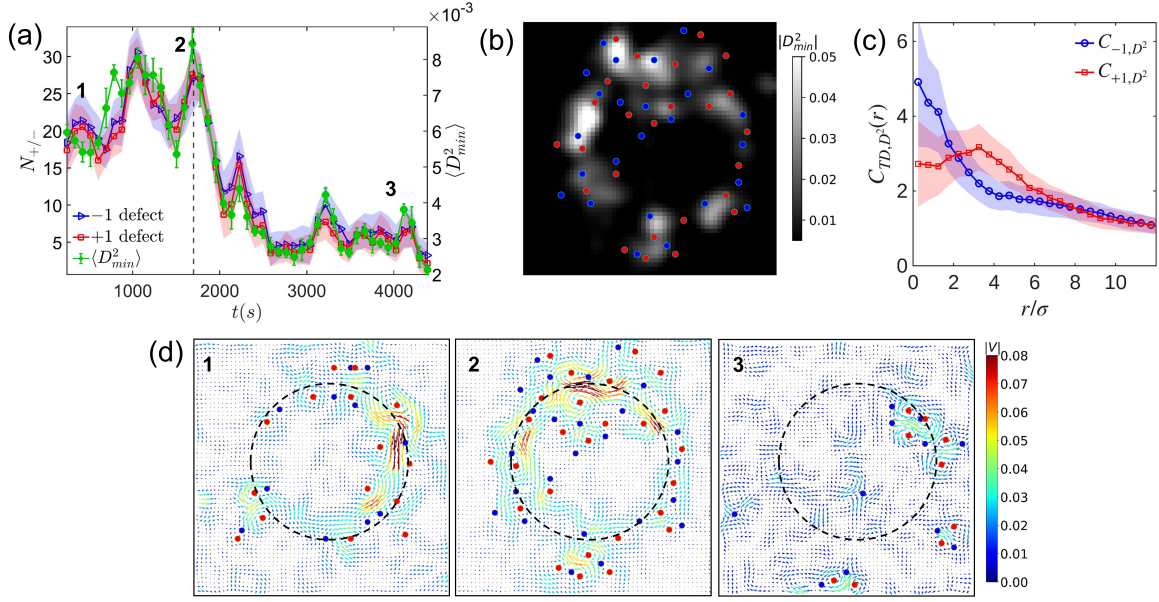}
\caption{Plastic deformation driven by topological defects during shear and subsequent relaxation following the cessation of shear.
(\textbf{a}) Temporal evolution of the number of positive and negative defects ($N_{+/-}$, left axis) and the non-affine displacement $D^2_{\mathrm{min}}$ (right axis). The vertical dashed line marks the cessation of shear. The numbers (1-3) point to different stages of deformation and relaxation, and the corresponding displacement fields are shown in panels (d1)–(d3). The error bars are obtained by averaging the data over a small time interval at any time $t$. 
(\textbf{b}) Topological defects overlaid on the coarse-grained non-affine displacement field $D^2$, color-coded in grayscale.
(\textbf{c}) Spatial correlation $C_{\mathrm{TD},D^2}$ between the locations of topological defects and coarse-grained sites with $D^2$ values in the top $5\%$ of the distribution. 
Background color in panels (a) and (c) represent error bar corresponding to standard deviations.
(\textbf{d}) Topological defects overlaid on the coarse-grained cage-relative displacement field. Positive ($q=+1$) and negative ($q=-1$) defects are shown as red and blue circles, respectively. Panels (d1) and (d2) correspond to sheared states at $\Theta = 0.05^{\circ}$ and $8.6^{\circ}$, respectively, while panel (d3) shows the configuration at $t=3672~\mathrm{s}$, after the laser is switched off. Only defects with large quadrupolar strength and circulation are displayed. Displacement vectors are color-coded according to their magnitude.}
\label{S32Fig2}
\end{figure*}
\subsection{Burgers Rings as Signatures of Eshelby Events}\label{sec2_sub6}

The studies discussed above demonstrate that topological defects provide valuable information about mechanically active regions and the collective organization of deformation in amorphous solids. An important remaining challenge, however, is the identification of a geometrical quantity capable of locating and characterizing the center of an elementary plastic event. In crystalline materials, Burgers vectors provide a natural measure of deformation incompatibility associated with dislocations. Extending this concept to amorphous solids remains nontrivial due to the absence of an underlying lattice.

A possible route was recently proposed by Bera \emph{et al.} \cite{bera2025burgersrings}, who introduced a continuous Burgers-vector field constructed from the coarse-grained displacement field. The approach associates a numerical local measure of incompatibility with each point in the continuum displacement field. This construction preserves the geometrical spirit of the classical Burgers circuit while remaining applicable to disordered materials \cite{liu2026}.

A key result of the study was that localized Eshelby-like rearrangements generate characteristic ring-shaped structures in magnitude of the Burgers-vector. Figure~\ref{sec2_fig6}(a) illustrates the emergence of this ``Burgers ring'' for numerical computation of multiple Eshelby inclusions organised horizontally. A separate Burgers rings appear for each of such events.  In Figure~\ref{sec2_fig6}(b) we have shown the displacemnt field for particle-based simulations. Here two Eshelby-like inclusions occurs, the magnitude of the Burgers field forms a closed ring surrounding the center of each of these events. 

An important advantage of this approach is that the ring structure offers a direct geometrical marker for locating plastic rearrangements. Burgers rings emerge naturally from the displacement field itself and remain clearly identifiable even when several nearby rearrangements coexist. This property suggests that they may provide a useful tool for detecting and tracking elementary plastic events in complex deformation patterns.

More broadly, the introduction of Burgers rings establishes a new connection between classical concepts originating in defect theory and modern topological descriptions of amorphous plasticity. Although the precise relationship between Burgers rings, topological charges, and plastic rearrangements remains an active topic of investigation, the results of Ref.~\cite{bera2025burgersrings} indicate that numerical Burgers field contains rich geometrical information that is not fully captured by conventional measures of plastic activity. These findings open new avenues for exploring the geometric measure of plastic events and their collective organization in disordered solids. 

\subsection{Generalization to Three Dimensions: Hedgehog Defects}\label{sec2_sub7}

Most of the studies discussed so far have focused on topological defects in two-dimensional vector fields, where the relevant excitations are vortices and anti-vortices characterized by an integer winding number. While these defects provide valuable insight into the topology of displacement and vibrational fields, they do not fully capture the richer topological structures that become possible in three dimensions. Extending topological descriptions of amorphous plasticity beyond two-dimensional projections is therefore an important step toward understanding realistic amorphous materials.

A natural generalization is provided by hedgehog defects, which correspond to point singularities in three-dimensional vector fields. As discussed in Sec.~\ref{section1}, these defects are characterized by the topological charge $Q$, defined in Eq.~\ref{eq_winding_3d}. The first systematic investigation of such defects in amorphous solids was recently reported by Bera \emph{et al.} \cite{bera2025hedgehog}. Using the three-dimensional non-affine displacement field generated during deformation, the authors identified point-like singularities, referred to as hedgehog topological defects (HTDs), and analyzed both their topological and geometrical properties.

In addition to the topological charge ($Q=\pm1$), the study introduced a geometrical classification of HTDs based on the local structure of the vector field. The defects were identified on a discrete cubic lattice \cite{kleman2003book}, and their nature was determined from the winding numbers measured on the six faces of the surrounding cubic cell. As illustrated in Fig.~\ref{sec2_fig4}(a), a radial defect yields a winding number $q=+1$ on all six faces, whereas a hyperbolic defect is characterized by two faces with $q=+1$ and four faces with $q=-1$. This construction provides a simple geometrical distinction between different classes of three-dimensional point defects.

A central result of the study is that the geometrical character of a defect carries significantly more information about local mechanical activity than its topological charge alone. When defects were classified solely according to $Q=\pm 1$, no difference was observed in their spatial correlation with soft spots. In contrast, a clear distinction emerged when radial and hyperbolic defects were considered separately. As shown in Fig.~\ref{sec2_fig4}(b), hyperbolic defects exhibit a substantially stronger correlation with soft spots than radial defects. Moreover, the correlation increases systematically with the degree of hyperbolicity, quantified by the parameter $N_s$, as illustrated in Fig.~\ref{sec2_fig4}(c). These observations suggest that local geometry, in addition to topology, plays an important role in determining the mechanical relevance of a defect.

The spatial relationship between HTDs and mechanically active regions is further illustrated in Fig.~\ref{sec2_fig4}(d), where the locations of defects are compared with regions of large non-affine displacement. A strong spatial overlap is evident. This trend is quantified in Figs.~\ref{sec2_fig4}(e,f). While defects with $Q=+1$ and $Q=-1$ display very similar correlations with soft spots, a slight distinction emerges when defects are classified according to their geometrical nature. HTDs when identified within non-affine field exhibit a stronger correlation with soft spots, but with minor differences in correlation between hyperbolic and radial nature.

The identification of hedgehog defects establishes a genuinely three-dimensional topological framework for analyzing deformation in amorphous solids. Whereas vortices and anti-vortices naturally arise in two-dimensional slices or projected vector fields, hedgehog defects provide direct access to the topology of the full three-dimensional displacement field. These developments considerably broaden the scope of topological approaches to glassy plasticity and suggest that the combined interplay of topology and geometry may provide a unified framework for understanding mechanical heterogeneity across different spatial dimensions.

\subsection{Geometrical and Topological Perspectives}\label{sec2_sub8}

The studies reviewed above show that topological defects can be identified in several mechanically relevant fields, including vibrational eigenmodes and non-affine displacement fields. These constructions are not identical, and they do not necessarily probe the same physical object. A central question is therefore how the different defect descriptions are related and which aspects of glassy plasticity each of them captures.

Recent work by Wu, Barrat, and Kob \cite{wu2026geometry} addressed the geometry of TDs associated with vibrational eigenmodes in a large three-dimensional glass. Their main finding is that, at low frequencies, these defects do not appear as isolated points. Instead, they form extended, quasi-linear filamentary structures. The number of defects scales as $\omega^2$ at low frequency. This geometry of TD line arrangements was argued to arise from the structure of low-frequency acoustic-like modes, whose wavelength controls the spacing between defect lines.

This result adds an important new perspective to the role of topology in amorphous solids. In two dimensions, topological defects in eigenvector fields appear as isolated point-like vortices and anti-vortices. In three dimensions, however, the zeros and phase singularities of these fields can organize into extended line-like networks. A particularly important finding is the connection between these defect networks and plastic deformation. Under athermal quasistatic shear, plastic events were found to correlate strongly with defects carrying negative topological charge. Moreover, the spatial correlations of plastic events exhibit a power-law decay with exponent $5/3$, closely matching the fractal organization of the defect-line network, characterized by a fractal dimension $d_f \approx d - 5/3$, with $d=3$. This correspondence suggests that plastic activity in three-dimensional glasses is not governed solely by independent local rearrangements, but is strongly influenced by the pre-existing geometrical organization of low-frequency vibrational eigenmodes.

This perspective helps clarify the relation among several approaches discussed in this review. Point defects in two-dimensional slices, Burgers rings around Eshelby-like events, hedgehog defects in three-dimensional displacement fields, and filamentary defects in vibrational eigenmodes highlight different aspects of the same broad problem: how disorder, elasticity, and topology organize the locations where plasticity occurs. At present, these descriptions are complementary rather than unified. Establishing their precise connections remains an important open challenge. The geometrical viewpoint emphasizes that topological defects in glasses should not be characterized only by their charge. Their spatial organization, connectivity, and scaling properties may be equally important for understanding yielding and energy dissipation. This provides a natural bridge from the simulation results discussed in this section to the experimental studies considered below.

\section{Topological Defects in Experimental Systems}\label{section3}
The studies reviewed in the previous section demonstrate that topological defects emerge naturally in a variety of vector fields associated with amorphous solids, including vibrational eigenmodes, non-affine displacement fields, and coarse-grained deformation fields. Most of these developments have been driven by numerical simulations and theoretical analyses, which provide direct access to particle trajectories, Hessian matrices, and displacement fields with high spatial resolution. Establishing the experimental relevance of these concepts, however, presents a considerably greater challenge.

Experimental detection generally relies on reconstructing appropriate dynamical or mechanical fields from particle-resolved measurements and subsequently extracting their topological properties. This requires a combination of high-resolution imaging, precise particle tracking, and suitable coarse-graining procedures capable of resolving the underlying vector fields. Recent advances in microscopy, confocal imaging, and particle-tracking techniques have made such analyses increasingly feasible, opening the possibility of studying topological defects directly in experimental amorphous materials.

Colloidal glasses provide a particularly attractive platform for these investigations. Owing to their accessible length and time scales, individual particle trajectories can be measured with high accuracy, allowing displacement fields, vibrational modes, and local rearrangements to be reconstructed at the single-particle level. As a result, colloidal systems have emerged as an important bridge between theoretical concepts and experimental observations of topological defects in disordered solids.

In this section, we review recent experimental developments in both two- and three-dimensional colloidal glasses. We begin with the first direct observation of topological defects in vibrational eigenmodes of a two-dimensional colloidal glass. We then discuss experiments in which shear is applied through an optical vortex, providing direct evidence for the role of topological defects during plastic deformation. Finally, we consider three-dimensional colloidal glasses under shear, where particle-resolved imaging enables the study of Eshelby-like rearrangements, yielding, and the emergence of three-dimensional topological defects. Together, these experiments provide growing evidence that topology is not merely a theoretical construct but constitutes an experimentally observable aspect of the mechanical response of amorphous solids.
\begin{figure*}[t]
\centering
\includegraphics[width=\textwidth]{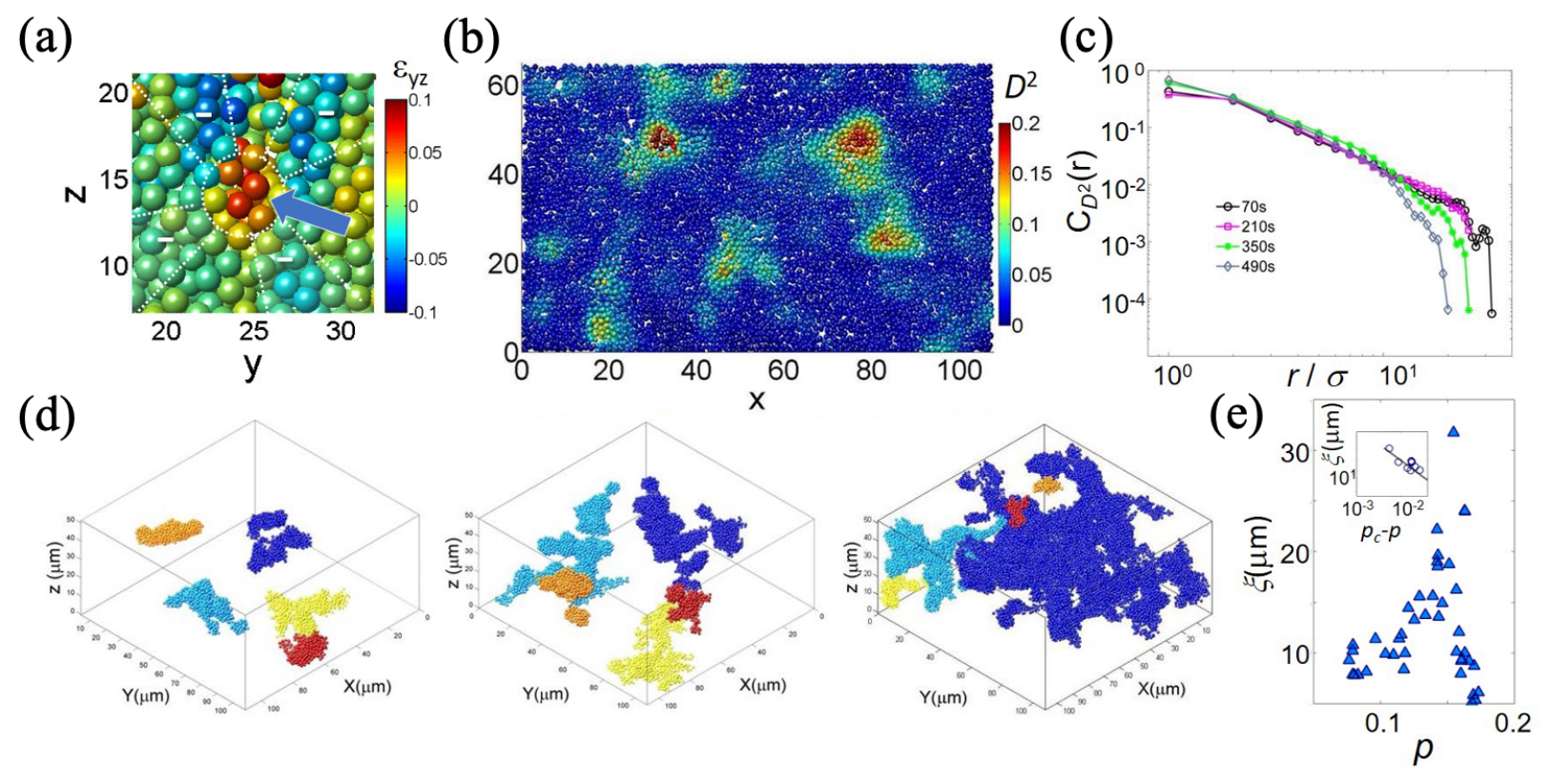}
\caption{Elements and correlations of plasticity in 3D sheared colloidal glasses. 
(\textbf{a}) Reconstructed shear strain reveals an Eshelby inclusion with surrounding quadrupolar strain field. 
(\textbf{b}) Distribution of nonaffine displacements in a  vertical cross section through a sheared colloidal glass (see color bar).
(\textbf{c}) Spatial correlations of non-affine displacements during the initial start-up shear approaching yielding. The correlations become exceeding long-range, with a power-law slope of $\sim -1.3$
(\textbf{d}) Reconstruction of particles with nonaffine displacements larger than the average. Highly nonaffine particles cluster in space, and the clusters grow with applied strain. Shown are reconstructions at applied shear strains 2.1, 4.9 and $10.1 \%$. 
(\textbf{e}) Correlation length of the largest cluster as a function of the fraction of highly nonaffine particles. Divergence is observed at a critical fraction $p_\mathrm{c} \sim 1.6$. Inset: scaling of the correlation length upon apporoach of $p_\mathrm{c}$. Panel (a) is adapted with permission from Schall \emph{et al.}, Science \textbf{318}, 1895 (2007) \cite{schall2007}. Copyright (2007) The American Association for the Advancement of Science. Panels (b-c) adapted with permission from Chikkadi \emph{et al.}, Phys. Rev. E \textbf{85}, 031402 (2012) \cite{Chikkadi12}. Copyright (2012) American Physical Society. Panels (d-e) adapted with permission from Ghosh \emph{et al.}, Phys. Rev. Lett. \textbf{118}, 148001 (2017) \cite{ghosh2017}. Copyright (2017) American Physical Society.}
%Panel (a) replotted from the original experimental data of Ref.~\cite{schall2007}. %if there is difficulty in getting permission from science.}
\label{Fig13}
\end{figure*}

\begin{figure*}[t]
\centering
\includegraphics[width=\textwidth]{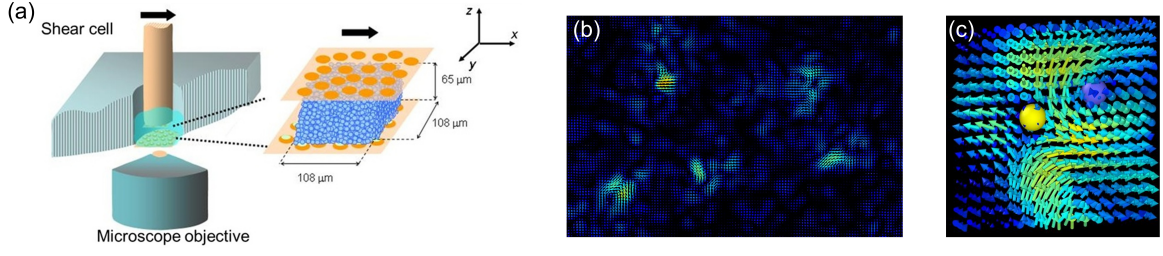}
\caption{Colloidal glasses under shear in 3D. 
(\textbf{a}) Schematic of the experimental setup showing a cross-section of shear-cell mounted on top of a confocal microscope. The shear-cell has two parallel boundaries, and the top boundary is moved relative to the bottom one. 
(\textbf{b}) A reconstruction of the coarse-grained nonaffine displacement field of particles in a 2D section.
(\textbf{c}) A positive and a negative hedgehog defects in a $2\sigma$ thick section along the $y-$ direction of the 3D data. Panel (a) adapted with permission from Chikkadi \emph{et al.}, Phys. Rev. E \textbf{85}, 031402 (2012) \cite{Chikkadi12}.
}
\label{S33Fig1}
\end{figure*}

\begin{figure*}[t]
\centering
\includegraphics[width=16cm]{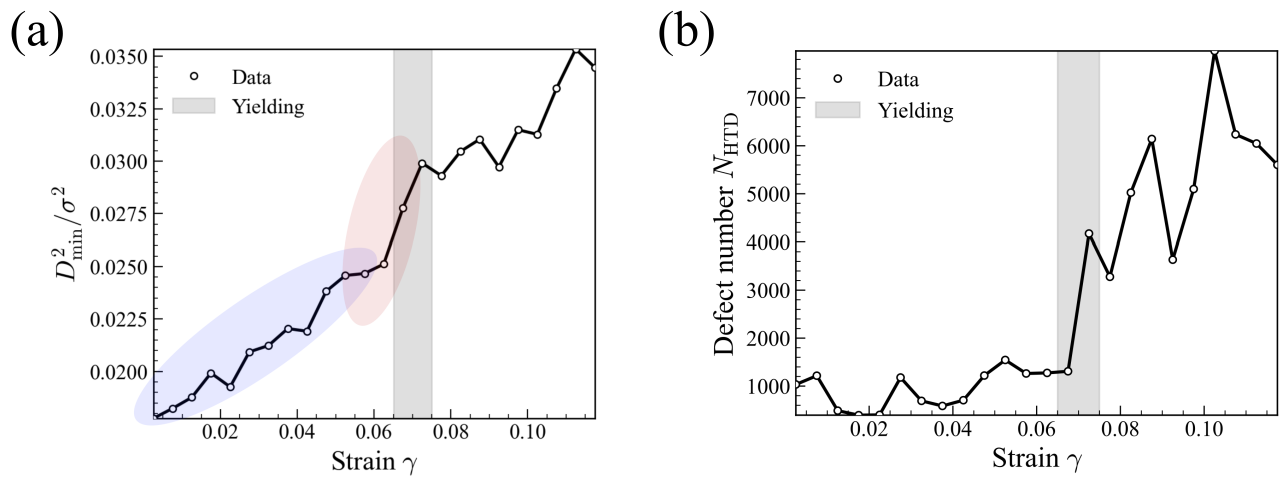}
\caption{Percolation of topological defects.
\textbf{(a)} Average non-affine displacement measure \(D^2_{\mathrm{min}}\) as a function of applied shear strain \(\gamma\). A change in slope is observed at the yielding transition, indicated by the shaded regions corresponding to the pre-yielding (blue) and yielding (red) regimes, reflecting the onset of enhanced non-affine motion.
\textbf{(b)} Total number of hedgehog topological defects, \(N_{\mathrm{HTD}}\), extracted from the three-dimensional non-affine displacement field as a function of \(\gamma\). In contrast to the gradual evolution of \(D^2_{\mathrm{min}}\), \(N_{\mathrm{HTD}}\) exhibits a pronounced spike at the yielding transition, providing a sharp topological signature of macroscopic plasticity.}
\label{S33Fig2}
\end{figure*}
%\color{red}{Here we need to briefly discuss the Matteo's experiment: In granular matter discuss \cite{wang2025topological}.}
%\color{black}
\subsection{Experimental Observation of Topological Defects in Colloidal Glasses}\label{sec3_sub1}

The first experimental identification of topological defects in a glassy system was achieved in a two-dimensional colloidal glass \cite{vaibhav2025}. The system consists of a bidisperse monolayer of paramagnetic colloidal particles confined at a water--air interface \cite{Keim17}. Since the particles interact through a well-characterized dipole--dipole repulsive potential, the Hessian matrix can be constructed directly from a static particle configuration. The behavior of the system is controlled by the dimensionless parameter $\Gamma = E_{\rm pot}/E_{\rm kin}$, which measures the ratio of potential to kinetic energy. As $\Gamma$ increases, the system evolves from a liquid-like state to a glassy state. In this study, $\Gamma = 423$, corresponding to a glassy state \cite{klix2015}. Diagonalization of the Hessian matrix constructed from the dipole--dipole interactions yields the vibrational eigenmodes of the experimentally realized amorphous solid.

Topological defects are subsequently identified from the eigenvector fields associated with these vibrational modes. After coarse-graining the local eigenvector orientations, vortices and anti-vortices are detected through the winding number of the orientation field. This procedure reveals quantized defects with charges $q=+1$ and $q=-1$, demonstrating that topological singularities are not merely numerical constructs but can be observed directly in an experimental glass.

The main results are summarized in Fig.~\ref{fig:colloidal_topodefects}. A schematic illustration of the colloidal monolayer is shown in Fig.~\ref{fig:colloidal_topodefects}(a). The vibrational density of states, $D(\omega)$, and the number of topological defects exhibit closely related behavior as functions of the eigenfrequency. Low-frequency modes display relatively simple and coherent defect patterns, whereas higher-frequency modes contain a denser and more disordered population of topological defects, as illustrated in Fig.~\ref{fig:colloidal_topodefects}(c,d).

Since the colloidal monolayer is not mechanically deformed in this experiment, soft spots cannot be identified through subsequent plastic rearrangements. Instead, they are determined directly from the vibrational eigenmodes using established mode-based approaches \cite{smessaert2014}. A particularly important observation is the strong spatial correlation between soft spots and topological defects. Defects carrying negative charge exhibit a significantly stronger correlation with soft regions at short distances than positively charged defects. This finding provides direct experimental evidence that the topology of vibrational eigenmodes encodes information about mechanical heterogeneity in amorphous solids and confirms the earlier numerical predictions of Wu \emph{et al.} \cite{wu2023}.

This work establishes an important experimental bridge to the simulation studies discussed earlier. It demonstrates that the topology of vibrational eigenmodes can be extracted from particle-resolved experimental data and that the resulting topological defects carry information about the spatial organization of soft regions, even in the absence of externally imposed deformation. The results therefore provide strong support for the emerging view that topological defects constitute a useful descriptor of mechanical heterogeneity in disordered solids.

%\begin{figure*}[t] \centering \includegraphics[width=\textwidth]{identification.png} \caption{ \textbf{Experimental identification of topological defects in a 2D colloidal glass.} \textbf{(Top left)} Schematic of the experimental setup: a bidisperse monolayer of superparamagnetic colloidal particles confined at a water--air interface and interacting via tunable dipole--dipole repulsion under an external magnetic field. \textbf{(Top right)} Vibrational density of states $D(\omega)$ (blue) and total number of topological defects $N_d$ (gray) as a function of frequency, showing a strong correlation between vibrational excitations and defect population. The inset highlights the linear scaling at low frequency. \textbf{(Bottom)} Eigenvector fields at low ($\omega=1.53$) and high ($\omega=141.15$) frequencies. Arrows indicate particle displacements, while red and blue dots denote vortices ($q=+1$) and anti-vortices ($q=-1$). Low-frequency modes exhibit coherent, extended swirling patterns with few defects, whereas high-frequency modes are increasingly disordered with a dense defect population. Adapted from Ref.~\cite{vaibhav2025}.} \label{fig:colloidal_topodefects} \end{figure*}

\subsection{Topological defects mediated plasticity of 2D colloidal glasses driven by an optical vortex}\label{sec3_sub2}
The topological defects associated with vibrational modes are now experimentally established in 2D quiescent colloidal glasses, however, their direct role in plastic deformation under shear has only recently begun to emerge. Recent experiments on two-dimensional amorphous granular packings under shear have provided further evidence for the role of topological defects in plasticity \cite{wang2025topological}. Topological defects identified in the displacement field were found to correlate strongly with plastic activity, while their collective alignment was shown to accompany the formation of shear bands. These results extend earlier simulation studies and support the idea that topological defects act as important carriers of plastic deformation in amorphous materials. Further experiments of two-dimensional colloidal glasses subjected to controlled shear using an optical vortex provide direct experimental evidence of the role of topological defects in plastic deformation \cite{sahoo2026}. These experiments based on particle-resolved microscopy and holographic optical tweezer enables direct tracking of microscopic dynamics during deformation. It employs an optical vortex generated using a spatial light modulator that imparts orbital angular momentum to colloidal particles confined within a ring geometry \cite{Grier03,Grier02}. The trapped particles experience a torque, thereby creating a shear zone in the surrounding glass. While the previous section described the identification of topological defects from vibrational eigenmodes derived from the Hessian matrix, the defects in the shear experiments are identified from coarse-grained cage-relative displacement fields. As discussed earlier, the winding number of the displacement orientation evaluated over closed loops determines the topological charge. This analysis reveals singular structures in the displacement field corresponding to vortices $q=+1$ and anti-vortices $q=-1$. Unlike crystalline dislocations, these defects emerge dynamically from heterogeneous nonaffine motion and remain robust in the presence of thermal fluctuations.

Figure~\ref{S32Fig1}(a) shows the experimental setup used to shear a colloidal monolayer using an optical vortex generated by a spatial light modulator. Particles trapped within the Laguerre–Gaussian beam acquire orbital angular momentum and rotate clockwise within a circular region of diameter $30\sigma$, transferring shear to the surrounding suspension and driving elastic-to-plastic deformation. The system is sheared by an angular displacement of $\Theta = 9.5^\circ$ over $\tau_{\textnormal{shear}} = 1700~\text{s}$, after which the laser is switched off and the suspension relaxes over $\tau_{\textnormal{relax}} \approx 3300~\text{s}$. Figure~\ref{S32Fig1}(b) shows particle trajectories superimposed on the reconstructed colloidal image, color-coded by integrated displacement, with the black circle indicating the beam region. The deformation induced by angular momentum transfer is characterized by the average angular displacement $\Theta$ of particles, computed in concentric circular bins around the vortex center as shown in Fig.~\ref{S32Fig1}(c). It reveals three distinct regions: a rigid core (red), a shear zone (blue) of width $\sim 16\sigma$ where the displacement decays rapidly with distance, and an outer region where the shear becomes negligible for $r>L/2$, with $L$ denoting the field of view size. The applied displacement $\Theta = 9.5^\circ$ is sufficiently large to drive yielding in the colloidal monolayer.

Topological defects are extracted from the displacement field using a two-step procedure: first computing the cage-relative (CR) displacements of particles and then coarse-graining the resulting field to identify topological singularities. The cage-relative displacement of a particle is defined as
\begin{equation}
\Delta \mathbf{r}_{CR}^i(t) = [\mathbf{r}^i(t) - \mathbf{r}^i(t-\Delta t)] - \frac{1}{N_i}\sum_{j=1}^{N_i}~ [\mathbf{r}^j(t) - \mathbf{r}^j(t-\Delta t)],
\end{equation}
where $N_i$ denotes the number of nearest neighbours in the first coordination shell. The displacements are evaluated over a fixed interval $\Delta t = \tau_{TD} = 200~\mathrm{s}$ to capture sufficient plastic activity. Subsequently, the cage-relative displacement field is coarse-grained onto a $60\sigma \times 60\sigma$ grid using a Gaussian weighting function, and topological defects are identified through the winding number, $q = \frac{1}{2\pi}\oint d\vartheta$, where $\vartheta$ is the phase of the coarse-grained field \cite{wu2023}. This procedure yields integer topological charges $q=-1,+1$, corresponding to antivortices, and vortices, respectively. Since thermal fluctuations can generate spurious defects that do not contribute to plasticity, we identify genuine defects based on their strength, quantified by the circulation for vortices and the quadrupolar amplitude for antivortices \cite{bera2025burgersrings}.

During shear, defect pairs nucleate and proliferate to accommodate stress through localized plastic rearrangements, whereas during relaxation the defect density decreases as oppositely charged defects annihilate. This can be inferred from the temporal evolution of the defect numbers ($N_{+/-}$) shown in Fig.~\ref{S32Fig2}(a) along with nonaffine displacement $D^2_{min}$, where the dashed line marks the cessation of shear. Figure~\ref{S32Fig2}(d) shows the strongest positive and negative topological defects, marked by red and blue circles, respectively, superimposed on the coarse-grained displacement field at different stages of deformation and relaxation. Panels (d1) and (d2) correspond to sheared states at $\Theta = 0.05^\circ$ and $\Theta = 8.6^\circ$, respectively, while panel (d3) shows the relaxed state after the shear is switched off.\\

Next, to directly correlate topological defects to plasticity, we compute the non-affine displacement $D^2_{\min}$, a standard measure of irreversible particle rearrangements \cite{falk1998}, over the same interval $\tau_{TD}=200~\mathrm{s}$ used for the displacement field. The average non-affine displacement, $\langle D^2_{\min} \rangle$, shown in Fig.~\ref{S32Fig2}(a), exhibits trends similar to the defect density $N_{+/-}$ during both shear and relaxation, suggesting a strong correlation between defect dynamics and plastic deformation. Figure~\ref{S32Fig2}(b) shows the coarse-grained non-affine displacement field, where bright regions correspond to enhanced plastic activity. Overlaying the topological defects reveals that they preferentially localize near these active regions. This correlation is quantified using the spatial correlation function \(C_{TD,D^2}(r)\), shown in Fig.~\ref{S32Fig2}(c), which demonstrates that negative defects exhibit a stronger and more localized correlation with plastic activity than positive defects, consistent with recent simulation studies \cite{wu2023}.\\

\subsection{Shear defects in 3D yielding colloidal glasses under shear: the Eshelby picture}\label{sec3_sub3}

Direct imaging of the particle-scale yielding of glasses under applied strain has been done on three-dimensional colloidal glasses under shear \cite{schall2007, chikkadi2011}. The explicit tracking of particle positions from 3D image stacks taken during an applied slow shear allows precise reconstruction of 3D particle trajectories. Earlier work has focused on the computation of  the strain fields and non-affine displacement field to identify plastic rearrangements (or so-called shear-transformation zones, STZ). The direct imaging revealed Eshelby-like inclusions~\cite{schall2007}, in which plastic rearrangements concentrate, and their surrounding quadrupolar strain field, as shown in Fig.~\ref{Fig13}(a). The characteristic long-range, quadrupolar strain field was observed to induce interactions between the Eshelby inclusions, triggering the formation of other aligned Eshelby inclusions in the vicinity, leading to emerging correlations and self organization~\cite{chikkadi2011}. The resulting field of coupled plastic rearrangements is shown in Fig.~\ref{Fig13}b, and the corresponding spatial correlations during startup shear in panel (c). The panel shows emerging long-range correlations with increasing shear strain approaching yielding. With increasing applied deformation, the non-affine displacement regions grow and become more numerous, until upon yielding, they percolate the system~\cite{ghosh2017}, as shown in panel (d). The percolation manifests in a large cluster of nonaffine displacements spreading across the system, visible as blue particle cluster in the last 3D snapshot in Fig.~\ref{Fig13}(d). The corresponding correlation length of the largest cluster is shown in Fig.~\ref{Fig13}(e); it diverges at a critical particle fraction of highly non-affine particles of $p_\mathrm{c} \sim 0.16$, as shown in the inset. At high shear rates, where the system cannot fully relax on the time scale dictated by the applied shear, the non-affine defects align along a plane parallel to the shear direction - resulting in the formation of a shear band~\cite{chikkadi2014}. Three-dimensional imaging during the initiation of a shear band clearly established the relation between the correlated alignment of Eshelby inclusions and the formation of the shear band. Overall, these results on the direct imaging of sheared colloidal glasses show that plasticity in amorphous solids is associated with the correlated formation and percolation of nonaffine rearrangements interacting via a Eshelby-like elastic strain fields.

\subsection{Percolation of topological defects at the yielding of 3D colloidal glasses under shear}\label{sec3_sub4}

In contrast to two-dimensional systems, topological defects in three dimensions possess a much richer structure, as discussed in Subsections \ref{sec1_sub6} and \ref{sec2_sub7} of this article. The deformation and displacement-related vector fields in disordered three-dimensional materials can support localized point singularities, referred to as hedgehog defects, emerging in eigenvector and non-affine displacement fields \cite{bera2025hedgehog}. In amorphous solids, these defects are found to preferentially occur in mechanically soft regions that are susceptible to plastic rearrangements and local instabilities, thereby offering a topological framework for understanding yielding transition in amorphous materials. Despite these advances, experimental demonstrations directly connecting these topological singularities with macroscopic yielding in three-dimensional amorphous materials are still largely lacking.

The experimental results obtained from the linear shear deformation of dense suspensions of sterically stabilized, fluorescently tagged polymethylmethacrylate (PMMA) colloids are outlined in this section. The particles are dispersed in a refractive-index- and density-matched solvent mixture, and the suspension is prepared near the colloidal glass transition \cite{Poon15, Weeks00}, where structural relaxation is strongly arrested, giving rise to a mechanically stable amorphous solid. The three-dimensional microstructure and dynamics are measured using confocal laser scanning microscopy, with 3D image stacks acquired during deformation to resolve particle positions with sub-pixel accuracy. Such volumetric stacks are acquired at regular intervals of strain steps, and the time-dependent particle trajectories are reconstructed using standard tracking algorithms. The applied shear is nearly quasi-static, achieved using a custom shear cell mounted on the microscope stage as shown in Fig.~\ref{S33Fig1}(a). The simultaneous shear and image acquisition enabled direct characterization of the instantaneous microstructure during deformation. 
This approach enables direct access to the instantaneous microstructure throughout the deformation process. The measured response displays an initial linear elastic regime at small $\gamma$, followed by a nonlinear yielding regime at larger strains. From the reconstructed trajectories, both affine and non-affine particle displacements are calculated between consecutive strain steps.

We follow a procedure similar to the one outlined in subsection~\ref{sec3_sub2} to determine the coarse-grained nonaffine displacements in 3D. The nonaffine displacements were obtained by subtracting the affine displacements of particles and further coarse-grained in 3D \cite{Chikkadi12}. A reconstruction of the coarse-grained non-affine displacement field in a 2D section in Fig.~\ref{S33Fig1}(b) reveals the presence of vortices and anti-vortices. We use the surface integral represented in the Eq.~\ref{eq_winding_3d} to determine the HTDs in the displacement field. A small section of the coarse-grained nonaffine displacement field shown in Fig.~\ref{S33Fig1}(c) reveals the presence of a positive and negative hedgehog topological defects in the vicinity of each other. 

A standard microscopic measure of irreversible deformation in amorphous solids is the non-affine displacement parameter $D_{\rm{min}}^{2}$, which characterizes the extent to which local particle motion deviates from an affine deformation within a local neighborhood. The averaged value of $D_{\rm{min}}^{2}$ therefore serves as a scalar indicator of the overall non-affine activity in the material. As illustrated in Fig.~\ref{S33Fig2}(a), the mean $D_{\rm{min}}^{2}$ grows continuously with increasing shear strain $\gamma$, with a noticeable change in slope appearing near the yielding regime. This increase reflects the progressive emergence of irreversible rearrangements and enhanced plastic dynamics. Although such behavior is broadly consistent with the conventional interpretation of yielding as a transition from predominantly elastic to plastic response, the variation of $D_{\rm{min}}^{2}$ remains smooth and does not by itself provide a distinct microscopic signature of yielding or reveal the collective organization underlying the transition.

Recent advances suggest that the topology singularities in the displacement fields may offer a more fundamental description of yielding phenomena in amorphous materials. In particular, numerical and theoretical studies have shown that three-dimensional non-affine displacement fields can host topological singularities that are strongly associated with mechanically unstable regions and plastic rearrangements. These observations suggest that yielding may involve a collective emergence of topology singularities in the displacement field itself. Inspired by this perspective, we investigate the topology of the experimentally measured three-dimensional non-affine displacement field in our colloidal glass system. By treating the local non-affine displacements as a coarse-grained vector field, we identify HTDs, which appear as point singularities carrying nonzero topological charge. The identification and characterization of these defects are performed directly using the particle-resolved experimental displacement data.

The topological analysis reveals a markedly different view of the yielding transition. Figure~\ref{S33Fig2}(b) shows that the total number of hedgehog defects, $N_{\rm{HTD}}$, remains relatively small within the elastic regime, but rises abruptly near the macroscopic yield point. Unlike the gradual increase observed in $D^{2}_{\rm{min}}$, the behavior of $N_{\rm{HTD}}$ is characterized by a pronounced peak concentrated around the yield strain. This feature is consistently reproduced across multiple experimental realizations and remains stable against reasonable changes in the defect-detection procedure. 

The rapid proliferation of hedgehog defects therefore provides a clear microscopic hallmark of yielding with an intrinsically topological origin. Rather than simply indicating an increase in local irreversible motion, the sharp rise in $N_{\rm{HTD}}$ reflects a collective restructuring of the non-affine displacement field, marked by the sudden appearance of topological singularities at the onset of plastic flow. The contrast between this singular topological response and the comparatively smooth behavior of conventional $D^2_{\rm{min}}$ suggests that yielding in three-dimensional amorphous solids is associated with a fundamental reorganization of the topology of the deformation field.

\section{Concluding Remarks}\label{section4}
The identification of topological defects in amorphous solids has opened a new avenue for understanding plasticity and mechanical failure in disordered materials. Recent studies have shown that well-defined topological structures emerge in vibrational eigenmodes, non-affine displacement fields and three-dimensional deformation fields. These defects exhibit strong correlations with soft regions, localized plastic rearrangements, yielding, and shear-band formation, and have now been observed in both simulations and experiments. Together, these developments suggest that topology provides a promising framework for connecting microscopic dynamics with macroscopic mechanical response in glasses.

\subsection{Advances and Shortcomings of Topological Methods}
One of the major advances of recent years has been the demonstration that amorphous solids can support robust topological structures despite the absence of long-range order. Vortices, anti-vortices, Burgers rings, and hedgehog defects provide geometrical and topological descriptors of mechanical heterogeneity across multiple length scales. These approaches have revealed strong links between topology and plastic activity, while also establishing connections between vibrational modes, non-affine elasticity, yielding, and strain localization. The recent experimental observation of topological defects in colloidal glasses \cite{vaibhav2025} and granular materials \cite{wang2025topological} further supports their physical relevance.

At the same time, a unified topological theory of amorphous plasticity remains elusive. Different approaches rely on different vector fields, coarse-graining procedures, and topological invariants, and the precise relationships among the resulting defect populations are not yet fully understood. Furthermore, although topological defects are often strongly correlated with plastic rearrangements, it remains unclear whether they constitute the fundamental carriers of plasticity or represent emergent signatures of underlying mechanical instabilities. Establishing these connections remains a central challenge for the field. In a broader context, conceptually related ideas have also been developed in topological superfluids, where quenched disorder gives rise to orbital, spin, and Weyl glass phases characterized by topologically non-trivial textures and defects rather than conventional crystalline order \cite{Volovik2019}. Although the underlying order parameters differ fundamentally from those of structural glasses, these studies illustrate how topology can provide a unifying language for describing disordered states across diverse condensed-matter systems. Likewise, continuum formulations based on gauge fields and tetrads have long emphasized the deep connections between defect topology and differential geometry \cite{Dzyaloshinskii1980}, suggesting promising directions for future continuum theories of amorphous plasticity.

\subsection{Open Questions}
Several important questions remain open. Can the various topological descriptions proposed so far be unified within a common theoretical framework? What is the precise relationship between topological defects, the energy landscape, and the microscopic mechanisms of plastic rearrangements? Can defect dynamics be used to predict yielding and failure before they occur? Addressing these questions will require closer integration of topology, continuum mechanics, statistical physics, machine learning, and experimental observations. Recent perspectives further suggest that future progress may depend on combining topological descriptors with geometrical, mechanical, and data-driven approaches to develop a more complete understanding of amorphous plasticity \cite{baggioli2026}. While a defect-based theory of glasses comparable to that available for crystalline materials has not yet emerged, recent progress strongly suggests that topology will play an increasingly important role in future studies of amorphous matter.

More broadly, the recent emergence of topological descriptions in glasses may signal a shift in how disorder is approached in condensed matter physics. Traditionally, the absence of translational order has been viewed as the primary obstacle preventing the identification of defects analogous to those found in crystals. The discovery that robust topological singularities arise naturally in vibrational, non-affine, and deformation-induced vector fields suggests an alternative viewpoint: while glasses may lack structural order, they can nevertheless possess well-defined topological organization in the fields that govern their mechanical response. In this sense, topology provides a language that is largely independent of microscopic structural details and may therefore offer a route toward a more universal description of amorphous matter. Whether topological defects ultimately play a role comparable to that of dislocations in crystals remains to be established, but the rapid convergence of theoretical predictions, numerical simulations, and experimental observations indicates that topology is likely to become an increasingly important component of the modern physics of glasses and disordered systems. The coming years will reveal whether these developments represent a new diagnostic framework for amorphous plasticity or the foundations of a genuinely unified defect theory for disordered solids.
An especially intriguing possibility is that topological observables may define universality classes of amorphous deformation that transcend microscopic details, allowing metallic glasses, colloidal glasses, granular materials, foams, and polymer glasses to be described within a common topological framework despite their vastly different microscopic interactions.

\section{Acknowledgements}
A.Z. gratefully acknowledges funding from the European Union through Horizon Europe ERC Grant No. 101043968 ``Multimech'' and from the Niedersächsische Akademie der Wissenschaften zu Göttingen in the frame of the Gauss Professorship program. A.Z. and A.B. gratefully acknowledge funding from U.S. Army Research Office through Contract No. W911NF-22-2-0256. V.C. acknowledges financial support from DST/SERB under the project grant CRG/2021/007824 and MHRD for Stars grant MoE-STARS/STARS-2/2023-0909. V.C. thanks Ratimanasee Sahu for help in preparing a few figures. P.S. acknowledges the funding from a VICI grant (No. 680.47.615) of the Netherlands Organization for Scinentific Research (NWO).

\section{Disclosure statement}
No potential conflict of interest was reported by the author(s).
%-----------------------------------------------------------------------------
%-------------------------------------------------------
% References
%-------------------------------------------------------

\printbibliography

@article{lerner2016_PRE,
  author = {Lerner, Edan},
  title = {Micromechanics of nonlinear plastic modes},
  journal = {Phys. Rev. E},
  volume = {93},
  number = {5},
  pages = {053004},
  year = {2016},
  doi = {10.1103/PhysRevE.93.053004}
}

@article{nicolas2018,
  author = {Nicolas, Alexandre and Ferrero, Ezequiel E. and Martens, Kirsten and Barrat, Jean-Louis},
  title = {Deformation and flow of amorphous solids: Insights from elastoplastic models},
  journal = {Rev. Mod. Phys.},
  volume = {90},
  number = {4},
  pages = {045006},
  year = {2018},
  doi = {10.1103/RevModPhys.90.045006}
}

@article{lin2015,
  author = {Lin, Jie and Gueudr{\'e}, Thomas and Rosso, Alberto and Wyart, Matthieu},
  title = {Criticality in the Approach to Failure in Amorphous Solids},
  journal = {Phys. Rev. Lett.},
  volume = {115},
  number = {16},
  pages = {168001},
  year = {2015},
  doi = {10.1103/PhysRevLett.115.168001}
}

@article{ferrero2021,
  author = {Ferrero, Ezequiel E. and Jagla, Eduardo A.},
  title = {Properties of the density of shear transformations in driven amorphous solids},
  journal = {J. Phys.: Condens. Matter},
  volume = {33},
  number = {12},
  pages = {124001},
  year = {2021},
  doi = {10.1088/1361-648X/abd73a}
}

@article{cubuk2015,
  author = {Cubuk, Ekin D. and Schoenholz, Samuel S. and Rieser, Jennifer M. and Malone, Brad D. and Rottler, J{\"o}rg and Durian, Douglas J. and Kaxiras, Efthimios and Liu, Andrea J.},
  title = {Identifying Structural Flow Defects in Disordered Solids Using Machine-Learning Methods},
  journal = {Phys. Rev. Lett.},
  volume = {114},
  number = {10},
  pages = {108001},
  year = {2015},
  doi = {10.1103/PhysRevLett.114.108001}
}

@article{bapst2020,
  author = {Bapst, Victor and Keck, T. and Grabska-Barwi{\'n}ska, Agnieszka and Donner, C. and Cubuk, Ekin D. and Schoenholz, Samuel S. and Obika, A. and Nelson, A. W. R. and Back, T. and Hassabis, Demis and Kohli, Pushmeet},
  title = {Unveiling the predictive power of static structure in glassy systems},
  journal = {Nat. Phys.},
  volume = {16},
  number = {4},
  pages = {448--454},
  year = {2020},
  doi = {10.1038/s41567-020-0842-8}
}

@article{boattini2020,
  author = {Boattini, Emanuele and Mar{\'i}n-Aguilar, Susana and Mitra, Saheli and Foffi, Giuseppe and Smallenburg, Frank and Filion, Laura},
  title = {Autonomously revealing hidden local structures in supercooled liquids},
  journal = {Nat. Commun.},
  volume = {11},
  number = {1},
  pages = {5479},
  year = {2020},
  doi = {10.1038/s41467-020-19286-8}
}

@article{moshe2015,
  author = {Moshe, Michael and Levin, Ido and Aharoni, Hillel and Kupferman, Raz and Sharon, Eran},
  title = {Geometry and mechanics of two-dimensional defects in amorphous materials},
  journal = {Proc. Natl. Acad. Sci. USA},
  volume = {112},
  number = {35},
  pages = {10873},
  year = {2015},
  doi = {10.1073/pnas.1506531112}
}

@article{yu2025,
  author = {Yu, Jinhua and Zhang, Zhen and Sha, Zhendong and Ding, Jun and Greer, A. Lindsay and Ma, Evan},
  title = {Structural state governs the mechanism of shear-band propagation in metallic glasses},
  journal = {Proc. Natl. Acad. Sci. USA},
  volume = {122},
  number = {27},
  pages = {e2427082122},
  year = {2025},
  doi = {10.1073/pnas.2427082122}
}

@article{sopu2017,
  author = {Şopu, Daniel and Stoica, Mihai and Eckert, Jürgen and Albe, Karsten},
  title = {Atomic-Level Processes of Shear Band Nucleation in Metallic Glasses},
  journal = {Phys. Rev. Lett.},
  volume = {119},
  number = {19},
  pages = {195503},
  year = {2017},
  doi = {10.1103/PhysRevLett.119.195503}
}

@article{hassani2019,
  author = {Hassani, Muhammad and Lagogianni, Alexandra E. and Varnik, Fathollah},
  title = {Probing the Degree of Heterogeneity within a Shear Band of a Model Glass},
  journal = {Phys. Rev. Lett.},
  volume = {123},
  number = {19},
  pages = {195502},
  year = {2019},
  doi = {10.1103/PhysRevLett.123.195502}
}

@article{Grier02,
  title={Dynamic holographic optical tweezers},
  author={Curtis, Jennifer E and Koss, Brian A and Grier, David G},
  journal={Opt. commun.},
  volume={207},
  number={1-6},
  pages={169--175},
  year={2002},
  publisher={Elsevier},
  doi ={10.1016/S0030-4018(02)01524-9}
}

@article{Grier03,
  title={Structure of optical vortices},
  author={Curtis, Jennifer E and Grier, David G},
  journal={Phys. Rev. Lett.},
  volume={90},
  number={13},
  pages={133901},
  year={2003},
  publisher={APS},
  doi ={10.1103/PhysRevLett.90.133901}
}

@article{Chikkadi12,
  title={Nonaffine measures of particle displacements in sheared colloidal glasses},
  author={Chikkadi, Vijayakumar and Schall, Peter},
  journal={Phys. Rev. E},
  volume={85},
  number={3},
  pages={031402},
  year={2012},
  publisher={APS},
  doi={10.1103/PhysRevE.85.031402}
}

@incollection{Poon15,
  title={Colloidal suspensions},
  author={Poon, Wilson CK},
  booktitle={The Oxford Handbook of Soft Condensed Matter},
  pages={1--50},
  year={2015},
  publisher={Oxford University Press Oxford}
}

@article{patinet2016,
  author  = {Patinet, Sylvain and Vandembroucq, Damien and Falk, Michael L.},
  title   = {Connecting Local Yield Stresses with Plastic Activity in Amorphous Solids},
  journal = {Phys. Rev. Lett.},
  volume  = {117},
  number  = {4},
  pages   = {045501},
  year    = {2016},
  doi     = {10.1103/PhysRevLett.117.045501}
}

@article{schoenholz2016,
  author  = {Schoenholz, Samuel S. and Cubuk, E. D. and Sussman, Daniel M. and Kaxiras, Efthimios and Liu, Andrea J.},
  title   = {A Structural Approach to Relaxation in Glassy Liquids},
  journal = {Nat. Phys.},
  volume  = {12},
  number  = {5},
  pages   = {469--471},
  year    = {2016},
  doi     = {10.1038/nphys3644}
}

@article{taylor1934,
  author = {Taylor, G. I.},
  title = {The Mechanism of Plastic Deformation of Crystals. Part I. Theoretical},
  journal = {Proc. R. Soc. A},
  volume = {145},
  number = {855},
  pages = {362--387},
  year = {1934},
  doi = {10.1098/rspa.1934.0106}
}

@article{orowan1934,
  author = {Orowan, E.},
  title = {Zur Kristallplastizität. III},
  journal = {Zeitschrift für Physik},
  volume = {89},
  number = {9--10},
  pages = {634--659},
  year = {1934},
  doi = {10.1007/BF01341480}
}

@article{polanyi1934,
  author = {Polanyi, M.},
  title = {Über eine Art Gitterstörung, die einen Kristall plastisch machen könnte},
  journal = {Zeitschrift für Physik},
  volume = {89},
  number = {9--10},
  pages = {660--664},
  year = {1934},
  doi = {10.1007/BF01341481}
}

@article{burgers1939,
  author = {Burgers, J. M.},
  title = {Some Considerations on the Fields of Stress Connected with Dislocations in a Regular Crystal Lattice},
  journal = {Proc. K. Ned. Akad. Wet.},
  volume = {42},
  pages = {293--325},
  year = {1939}
}

@article{frank1951,
  author = {Frank, F. C.},
  title = {The Resultant Content of Dislocations in an Arbitrary Intercrystalline Boundary},
  journal = {Symp. Plast. Deform. Cryst. Solids},
  pages = {150--154},
  year = {1951}
}

@article{nabarro1947,
  author = {Nabarro, F. R. N.},
  title = {Dislocations in a Simple Cubic Lattice},
  journal = {Proc. Phys. Soc.},
  volume = {59},
  number = {2},
  pages = {256--272},
  year = {1947},
  doi = {10.1088/0959-5309/59/2/309}
}

@book{hirth1982,
  author = {Hirth, John P. and Lothe, Jens},
  title = {Theory of Dislocations},
  edition = {2nd},
  publisher = {Wiley},
  address = {New York},
  year = {1982}
}

@book{mura1987,
  author = {Mura, Toshio},
  title = {Micromechanics of Defects in Solids},
  edition = {2nd},
  publisher = {Martinus Nijhoff},
  address = {Dordrecht},
  year = {1987},
  doi = {10.1007/978-94-009-3489-4}
}

@article{volterra1907,
  author = {Volterra, V.},
  title = {Sur l'équilibre des corps élastiques multiplement connexes},
  journal = {Annales Scientifiques de l'École Normale Supérieure},
  volume = {24},
  pages = {401--517},
  year = {1907},
  doi = {10.24033/asens.583}  
}

@article{frank1958,
  author = {Frank, F. C.},
  title = {On the Theory of Liquid Crystals},
  journal = {Discussions of the Faraday Society},
  volume = {25},
  pages = {19--28},
  year = {1958},
  doi = {10.1039/DF9582500019}
}

@book{kleman1977,
  author = {Kléman, Maurice},
  title = {Points, Lines and Walls: In Liquid Crystals, Magnetic Systems and Various Ordered Media},
  publisher = {Wiley},
  address = {New York},
  year = {1977}
}

@article{kleman2008,
  author = {Kléman, Maurice and Friedel, Jacques},
  title = {Disclinations, Dislocations, and Continuous Defects: A Reappraisal},
  journal = {Rev. Mod. Phys.},
  volume = {80},
  number = {1},
  pages = {61--115},
  year = {2008},
  doi = {10.1103/RevModPhys.80.61}
}

@article{steinhardt1981_1,
  author        = {Steinhardt, P. J. and Nelson, D. R. and Ronchetti, M.},
  title         = {Icosahedral Bond Orientational Order in Supercooled Liquids},
  journal       = {Phys. Rev. Lett.},
  volume        = {47},
  number        = {18},
  pages         = {1297--1300},
  year          = {1981},
  doi           = {10.1103/PhysRevLett.47.1297},
  url           = {https://doi.org/10.1103/PhysRevLett.47.1297},
}

@article{steinhardt1981_2,
  author        = {Steinhardt, P. J. and Chaudhari, P.},
  title         = {Point and line defects in glasses},
  journal       = {Philos. Mag. A},
  volume        = {44},
  number        = {6},
  pages         = {1375--1381},
  year          = {1981},
  doi           = {10.1080/01418618108235816},
  url           = {https://doi.org/10.1080/01418618108235816},
}

@article{steinhardt1983,
  author        = {Steinhardt, P. J. and Nelson, D. R. and Ronchetti, M.},
  title         = {Bond-orientational order in liquids and glasses},
  journal       = {Phys. Rev. B},
  volume        = {28},
  number        = {2},
  pages         = {784--805},
  year          = {1983},
  doi           = {10.1103/PhysRevB.28.784},
  url           = {https://doi.org/10.1103/PhysRevB.28.784},
}

@article{nelson1983,
  author        = {Nelson, D. R.},
  title         = {Order, frustration, and defects in liquids and glasses},
  journal       = {Phys. Rev. B},
  volume        = {28},
  number        = {10},
  pages         = {5515--5535},
  year          = {1983},
  doi           = {10.1103/PhysRevB.28.5515},
  url           = {https://doi.org/10.1103/PhysRevB.28.5515},
}

@article{anderson1995,
  author = {Anderson, P. W.},
  title = {Through the Glass Lightly},
  journal = {Science},
  volume = {267},
  number = {5204},
  pages = {1615--1616},
  year = {1995},
  doi = {10.1126/science.267.5204.1615.f},
  url={https://www.science.org/doi/10.1126/science.267.5204.1615.f},
  
}

@article{angell1995,
  author = {Angell, C. A.},
  title = {Formation of Glasses from Liquids and Biopolymers},
  journal = {Science},
  volume = {267},
  number = {5206},
  pages = {1924--1935},
  year = {1995},
  doi = {10.1126/science.267.5206.1924}
}

@article{debenedetti2001,
  author = {Debenedetti, P. G. and Stillinger, F. H.},
  title = {Supercooled Liquids and the Glass Transition},
  journal = {Nature},
  volume = {410},
  pages = {259--267},
  year = {2001},
  doi = {10.1038/35065704}
}

@article{berthier2011,
  author = {Berthier, L. and Biroli, G.},
  title = {Theoretical Perspective on the Glass Transition and Amorphous Materials},
  journal = {Rev. Mod. Phys.},
  volume = {83},
  pages = {587--645},
  year = {2011},
  doi = {10.1103/RevModPhys.83.587}
}

@article{ediger2000,
  author = {Ediger, M. D.},
  title = {Spatially Heterogeneous Dynamics in Supercooled Liquids},
  journal = {Annu. Rev. Phys. Chem.},
  volume = {51},
  pages = {99--128},
  year = {2000},
  doi = {10.1146/annurev.physchem.51.1.99}
}

@article{tanaka2010,
  author = {Tanaka, Hajime and Kawasaki, Takeshi and Shintani, Hiroshi and Watanabe, Koji},
  title = {Critical-like Behaviour of Glass-Forming Liquids},
  journal = {Nat. Mater.},
  volume = {9},
  pages = {324--331},
  year = {2010},
  doi = {10.1038/nmat2634}
}

@article{tanaka2019,
  author = {Tanaka, Hajime and Tong, Hua and Shi, Runhua and Russo, John},
  title = {Revealing Key Structural Features Hidden in Liquids and Glasses},
  journal = {Nat. Rev. Phys.},
  volume = {1},
  pages = {333--348},
  year = {2019},
  doi = {10.1038/s42254-019-0053-3}
}

@article{argon1979,
  author = {Argon, A. S.},
  title = {Plastic Deformation in Metallic Glasses},
  journal = {Acta Metallurgica},
  volume = {27},
  number = {1},
  pages = {47--58},
  year = {1979},
  doi = {10.1016/0001-6160(79)90055-5}
}

@article{falk1998,
  author = {Falk, Michael L. and Langer, J. S.},
  title = {Dynamics of Viscoplastic Deformation in Amorphous Solids},
  journal = {Physical Review E},
  volume = {57},
  pages = {7192--7205},
  year = {1998},
  doi = {10.1103/PhysRevE.57.7192}
}

@article{Weeks00,
  title={Three-dimensional direct imaging of structural relaxation near the colloidal glass transition},
  author={Weeks, Eric R and Crocker, John C and Levitt, Andrew C and Schofield, Andrew and Weitz, David A},
  journal={Science},
  volume={287},
  number={5453},
  pages={627--631},
  year={2000},
  publisher={American Association for the Advancement of Science},
  doi ={10.1126/science.287.5453.627}
}

@article{cohen1959,
  author = {Cohen, Morris H. and Turnbull, David},
  title = {Molecular Transport in Liquids and Glasses},
  journal = {J. Chem. Phys.},
  volume = {31},
  number = {5},
  pages = {1164--1169},
  year = {1959},
  doi = {10.1063/1.1730566}
}

@article{turnbull1961,
  author = {Turnbull, David and Cohen, Morris H.},
  title = {Free-Volume Model of the Amorphous Phase: Glass Transition},
  journal = {J. Chem. Phys.},
  volume = {34},
  number = {1},
  pages = {120--125},
  year = {1961},
  doi = {10.1063/1.1731549}
}

@article{spaepen1977,
  author = {Spaepen, Frans},
  title = {A Microscopic Mechanism for Steady State Inhomogeneous Flow in Metallic Glasses},
  journal = {Acta Metall.},
  volume = {25},
  number = {4},
  pages = {407--415},
  year = {1977},
  doi = {10.1016/0001-6160(77)90232-2}
}

@article{langer2008,
  author = {Langer, J. S.},
  title = {Shear-Transformation-Zone Theory of Plastic Deformation Near the Glass Transition},
  journal = {Phys. Rev. E},
  volume = {77},
  number = {2},
  pages = {021502},
  year = {2008},
  doi = {10.1103/PhysRevE.77.021502}
}

@article{manning2007,
  author = {Manning, M. L. and Langer, J. S. and Carlson, J. M.},
  title = {Strain Localization in a Shear Transformation Zone Model for Amorphous Solids},
  journal = {Physical Review E},
  volume = {76},
  number = {5},
  pages = {056106},
  year = {2007},
  doi = {10.1103/PhysRevE.76.056106}
}

@article{greer1995,
  author = {Greer, A. Lindsay},
  title = {Metallic Glasses},
  journal = {Science},
  volume = {267},
  number = {5206},
  pages = {1947--1953},
  year = {1995},
  doi = {10.1126/science.267.5206.1947}
}

@article{schuh2007,
  author = {Schuh, Christopher A. and Hufnagel, Todd C. and Ramamurty, U.},
  title = {Mechanical Behavior of Amorphous Alloys},
  journal = {Acta Mater.},
  volume = {55},
  number = {12},
  pages = {4067--4109},
  year = {2007},
  doi = {10.1016/j.actamat.2007.01.052}
}

@article{wang2012,
  author = {Wang, Wei H.},
  title = {The Elastic Properties, Elastic Models and Elastic Perspectives of Metallic Glasses},
  journal = {Prog. Mater. Sci.},
  volume = {57},
  number = {3},
  pages = {487--656},
  year = {2012},
  doi = {10.1016/j.pmatsci.2011.07.001}
}

@article{johnson2005,
  author = {Johnson, W. L. and Samwer, K.},
  title = {A Universal Criterion for Plastic Yielding of Metallic Glasses with a $(T/T_g)^{2/3}$ Temperature Dependence},
  journal = {Phys. Rev. Lett.},
  volume = {95},
  pages = {195501},
  year = {2005},
  doi = {10.1103/PhysRevLett.95.195501}
}

@article{wang2015,
  author = {Wang, Wei H.},
  title = {Dynamic relaxations and relaxation-property relationships in metallic glasses},
  journal = {Prog. Mater. Sci.},
  volume = {106},
  pages = {100561},
  year = {2019},
  doi = {10.1016/j.pmatsci.2019.03.006}
}

@article{frank1952,
  author = {Frank, F. C.},
  title = {Supercooling of Liquids},
  journal = {Proc. R. Soc. A},
  volume = {215},
  number = {1120},
  pages = {43--46},
  year = {1952},
  doi = {10.1098/rspa.1952.0194}
}

@article{nelson1985,
  author = {Nelson, David R.},
  title = {Liquids and Glasses in Spaces of Incommensurate Curvature},
  journal = {Phys. Rev. Lett.},
  volume = {50},
  pages = {982--985},
  year = {1983},
  doi = {10.1103/PhysRevLett.50.982}
}

@article{kivelson1995,
  author = {Kivelson, D. and Kivelson, S. A. and Zhao, X. L. and Nussinov, Z. and Tarjus, G.},
  title = {A Thermodynamic Theory of Supercooled Liquids},
  journal = {Physica A},
  volume = {219},
  number = {1--2},
  pages = {27--38},
  year = {1995},
  doi = {10.1016/0378-4371(95)00140-3}
}

@article{tarjus1995,
  author = {Tarjus, Gilles and Kivelson, David},
  title = {Breakdown of the Stokes-Einstein Relation in Supercooled Liquids},
  journal = {J. Chem. Phys.},
  volume = {103},
  pages = {3071--3073},
  year = {1995},
  doi = {10.1063/1.470495}
}

@article{tarjus2005,
  author = {Tarjus, Gilles and Kivelson, David and Nussinov, Zohar and Viot, Pascal},
  title = {The Frustration-Based Approach of Supercooled Liquids and the Glass Transition: A Review and Critical Assessment},
  journal = {J. Phys.: Condens. Matter},
  volume = {17},
  number = {50},
  pages = {R1143--R1182},
  year = {2005},
  doi = {10.1088/0953-8984/17/50/R01}
}

@article{miracle2004,
  author = {Miracle, Daniel B.},
  title = {A Structural Model for Metallic Glasses},
  journal = {Nat. Mater.},
  volume = {3},
  pages = {697--702},
  year = {2004},
  doi = {10.1038/nmat1219}
}

@article{sheng2006,
  author = {Sheng, H. W. and Luo, W. K. and Alamgir, F. M. and Bai, J. M. and Ma, E.},
  title = {Atomic Packing and Short-to-Medium-Range Order in Metallic Glasses},
  journal = {Nature},
  volume = {439},
  pages = {419--425},
  year = {2006},
  doi = {10.1038/nature04421}
}

@article{hirata2013,
  author = {Hirata, A. and Guan, P. and Fujita, T. and Hirotsu, Y. and Inoue, A. and Yavari, A. R. and Sakurai, T. and Chen, M. W.},
  title = {Direct observation of local atomic order in a metallic glass},
  journal = {Nat. Mater.},
  volume = {10},
  pages = {28--33},
  year = {2011},
  doi = {10.1038/nmat2897}
}

@article{coslovich2011,
  author = {Coslovich, Daniele},
  title = {Locally preferred structures and many-body static correlations in viscous liquids},
  journal = {Phys. Rev. E},
  volume = {83},
  pages = {051505},
  year = {2011},
  doi = {10.1103/PhysRevE.83.051505}
}

@article{royall2015,
  author = {Royall, C. Patrick and Williams, Stephen R.},
  title = {The Role of Local Structure in Dynamical Arrest},
  journal = {Phys. Rep.},
  volume = {560},
  pages = {1--75},
  year = {2015},
  doi = {10.1016/j.physrep.2014.11.001}
}

@article{eshelby1957,
  author = {Eshelby, J. D.},
  title = {The Determination of the Elastic Field of an Ellipsoidal Inclusion, and Related Problems},
  journal = {Proc. R. Soc. A},
  volume = {241},
  number = {1226},
  pages = {376--396},
  year = {1957},
  doi = {10.1098/rspa.1957.0133}
}

@article{eshelby1959,
  author = {Eshelby, J. D.},
  title = {The Elastic Field Outside an Ellipsoidal Inclusion},
  journal = {Proc. R. Soc. A},
  volume = {252},
  number = {1271},
  pages = {561--569},
  year = {1959},
  doi = {10.1098/rspa.1959.0173}
}

@article{picard2004,
  author = {Picard, Gilles and Ajdari, Armand and Bocquet, Lyderic and Lequeux, François},
  title = {Elastic Consequences of a Single Plastic Event: A Step Towards the Microscopic Modeling of the Flow of Yield Stress Fluids},
  journal = {Eur. Phys. J. E},
  volume = {15},
  number = {4},
  pages = {371--381},
  year = {2004},
  doi = {10.1140/epje/i2004-10054-8}
}

@article{maloney2004,
  author = {Maloney, Craig E. and Lemaître, Anaël},
  title = {Subextensive Scaling in the Athermal, Quasistatic Limit of Amorphous Matter in Plastic Shear Flow},
  journal = {Phys. Rev. Lett.},
  volume = {93},
  pages = {016001},
  year = {2004},
  doi = {10.1103/PhysRevLett.93.016001}
}

@article{maloney2006,
  author = {Maloney, Craig E. and Lemaître, Anaël},
  title = {Amorphous Systems in Athermal, Quasistatic Shear},
  journal = {Phys. Rev. E},
  volume = {74},
  pages = {016118},
  year = {2006},
  doi = {10.1103/PhysRevE.74.016118}
}

@article{tanguy2006,
  author = {Tanguy, Anne and Leonforte, Fabrice and Barrat, Jean-Louis},
  title = {Plastic Response of a 2D Lennard-Jones Amorphous Solid: Detailed Analysis of the Local Rearrangements at Very Slow Strain Rate},
  journal = {Eur. Phys. J. E},
  volume = {20},
  number = {3},
  pages = {355--364},
  year = {2006},
  doi = {10.1140/epje/i2006-10024-2}
}

@article{lerner2009,
  author = {Lerner, Edan and Procaccia, Itamar},
  title = {Locality and Nonlocality in Elastoplastic Responses of Amorphous Solids},
  journal = {Phys. Rev. E},
  volume = {79},
  pages = {066109},
  year = {2009},
  doi = {10.1103/PhysRevE.79.066109}
}

@article{dasgupta2013,
  author = {Dasgupta, Ratul and Hentschel, H. G. E. and Procaccia, Itamar},
  title = {Yield Strain in Shear Banding Amorphous Solids},
  journal = {Phys. Rev. Lett.},
  volume = {109},
  pages = {255502},
  year = {2012},
  doi = {10.1103/PhysRevLett.109.255502}
}

@article{hentschel2015,
  author = {Hentschel, H. G. E. and Karmakar, Smarajit and Lerner, Edan and Procaccia, Itamar},
  title = {Do Athermal Amorphous Solids Exist?},
  journal = {Phys. Rev. E},
  volume = {83},
  pages = {061101},
  year = {2011},
  doi = {10.1103/PhysRevE.83.061101}
}

@article{bouchbinder2007,
  author = {Bouchbinder, Eran and Langer, J. S.},
  title = {Nonequilibrium Thermodynamics of Driven Amorphous Materials. I. Internal Degrees of Freedom and Volume Deformation},
  journal = {Phys. Rev. E},
  volume = {80},
  pages = {031131},
  year = {2009},
  doi = {10.1103/PhysRevE.80.031131}
}

@article{sengupta2012,
  author = {Dasgupta, R. and Gendelman, O. and Mishra, P. and Procaccia, I. and Shor, C. A. B. Z. },
  title = {Shear localization in three-dimensional amorphous solids},
  journal = {Phys. Rev. E},
  volume = {88},
  pages = {032401},
  year = {2013},
  doi = {10.1103/PhysRevE.88.032401},
}

@article{alexander1998,
  author = {Alexander, S.},
  title = {Amorphous Solids: Their Structure, Lattice Dynamics and Elasticity},
  journal = {Phys. Rep.},
  volume = {296},
  number = {2--4},
  pages = {65--236},
  year = {1998},
  doi = {10.1016/S0370-1573(97)00069-0}
}

@article{lemaitre2006,
  author = {Lemaître, Anaël and Maloney, Craig E.},
  title = {Sum Rules for the Quasi-static and Visco-elastic Response of Disordered Solids at Zero Temperature},
  journal = {J. Stat. Phys.},
  volume = {123},
  number = {2},
  pages = {415--453},
  year = {2006},
  doi = {10.1007/s10955-005-9015-5}
}

@article{zaccone2011,
  author = {Zaccone, Alessio and Scossa-Romano, Emanuela},
  title = {Approximate Analytical Description of the Nonaffine Response of Amorphous Solids},
  journal = {Phys. Rev. B},
  volume = {83},
  pages = {184205},
  year = {2011},
  doi = {10.1103/PhysRevB.83.184205}
}

@article{eugene1986,
  author = {Lavrentovich, O. D. and Terent'ev, E. M.},
  title = {Phase transition altering the symmetry of topological point defects (hedgehogs) in a nematic liquid crystal},
  journal = {Soviet Physics JETP},
  volume = {64},
  number = {6},
  pages = {1237--1244},
  year = {1986}
}

@article{zaccone2013,
  author = {Zaccone, Alessio and Terentjev, Eugene M.},
  title = {Disorder-Assisted Melting and the Glass Transition in Amorphous Solids},
  journal = {Phys. Rev. Lett.},
  volume = {110},
  pages = {178002},
  year = {2013},
  doi = {10.1103/PhysRevLett.110.178002}
}

@book{zaccone2023book,
  author = {Zaccone, Alessio},
  title = {Theory of Disordered Solids},
  publisher = {Springer},
  address = {Cham},
  year = {2022},
  doi = {10.1007/978-3-031-24706-4}
}

@article{tanguy2002,
  author = {Tanguy, Anne and Wittmer, J. P. and Leonforte, Fabrice and Barrat, Jean-Louis},
  title = {Continuum Limit of Amorphous Elastic Bodies: A Finite-Size Study of Low-Frequency Harmonic Vibrations},
  journal = {Phys. Rev. B},
  volume = {66},
  pages = {174205},
  year = {2002},
  doi = {10.1103/PhysRevB.66.174205}
}

@article{leonforte2005,
  author = {Leonforte, Fabrice and Boissière, Romain and Tanguy, Anne and Wittmer, J. P. and Barrat, Jean-Louis},
  title = {Continuum Limit of Amorphous Elastic Bodies. III. Three-Dimensional Systems},
  journal = {Phys. Rev. B},
  volume = {72},
  pages = {224206},
  year = {2005},
  doi = {10.1103/PhysRevB.72.224206}
}

@article{didonna2005,
  author = {DiDonna, B. A. and Lubensky, T. C.},
  title = {Nonaffine Correlations in Random Elastic Media},
  journal = {Phys. Rev. E},
  volume = {72},
  pages = {066619},
  year = {2005},
  doi = {10.1103/PhysRevE.72.066619}
}

@article{ellenbroek2006,
  author = {Ellenbroek, Wouter G. and Zeravcic, Zorana and van Saarloos, Wim and van Hecke, Martin},
  title = {Non-affine Response: Jammed Packings vs. Spring Networks},
  journal = {Europhys. Lett.},
  volume = {87},
  pages = {34004},
  year = {2009},
  doi = {10.1209/0295-5075/87/34004}
}

@article{wyart2005,
  author = {Wyart, Matthieu and Nagel, Sidney R. and Witten, Thomas A.},
  title = {Geometric Origin of Excess Low-Frequency Vibrational Modes in Weakly Connected Amorphous Solids},
  journal = {Europhys. Lett.},
  volume = {72},
  number = {3},
  pages = {486--492},
  year = {2005},
  doi = {10.1209/epl/i2005-10245-5}
}

@article{wyart2010,
  author = {Wyart, Matthieu},
  title = {Scaling of Phononic Transport with Connectivity in Amorphous Solids},
  journal = {Europhys. Lett.},
  volume = {89},
  pages = {64001},
  year = {2010},
  doi = {10.1209/0295-5075/89/64001}
}

@article{lerner2014,
  author = {Lerner, Edan and Düring, Gustavo and Wyart, Matthieu},
  title = {Low-Energy Nonlinear Excitations in Sphere Packings},
  journal = {Soft Matter},
  volume = {9},
  pages = {8252--8263},
  year = {2013},
  doi = {10.1039/C3SM50515D}
}

@article{manning2011,
  author = {Manning, M. Lisa and Liu, Andrea J.},
  title = {Vibrational Modes Identify Soft Spots in a Sheared Disordered Packing},
  journal = {Phys. Rev. Lett.},
  volume = {107},
  pages = {108302},
  year = {2011},
  doi = {10.1103/PhysRevLett.107.108302}
}

@article{Zaccone2022,
  title = {Explicit Analytical Solution for Random Close Packing in $d=2$ and $d=3$},
  author = {Zaccone, Alessio},
  journal = {Phys. Rev. Lett.},
  volume = {128},
  issue = {2},
  pages = {028002},
  numpages = {5},
  year = {2022},
  publisher = {American Physical Society},
  doi = {10.1103/PhysRevLett.128.028002},
  url = {https://link.aps.org/doi/10.1103/PhysRevLett.128.028002}
}

@article{Truskett,
  title = {Is Random Close Packing of Spheres Well Defined?},
  author = {Torquato, S. and Truskett, T. M. and Debenedetti, P. G.},
  journal = {Phys. Rev. Lett.},
  volume = {84},
  issue = {10},
  pages = {2064--2067},
  numpages = {0},
  year = {2000},
  publisher = {American Physical Society},
  doi = {10.1103/PhysRevLett.84.2064},
  url = {https://link.aps.org/doi/10.1103/PhysRevLett.84.2064}
}

@article{willmarth2025,
  title = {Particle-scale origin of quadrupolar nonaffine displacement fields in granular solids},
  author = {Willmarth, Evan P. and Jin, Weiwei and Wang, Dong and Datye, Amit and Schwarz, Udo D. and Shattuck, Mark D. and O'Hern, Corey S.},
  journal = {Phys. Rev. E},
  volume = {112},
  issue = {5},
  pages = {055402},
  numpages = {21},
  year = {2025},
  publisher = {American Physical Society},
  doi = {10.1103/42dk-54hl},
  url = {https://link.aps.org/doi/10.1103/42dk-54hl}
}

@article{zaccone2025jamming,
    author = {Zaccone, Alessio},
    title = {Complete mathematical theory of the jamming transition: A perspective},
    journal = {Journal of Applied Physics},
    volume = {137},
    number = {5},
    pages = {050901},
    year = {2025},
    month = {02},
    issn = {0021-8979},
    doi = {10.1063/5.0245684},
    url = {https://doi.org/10.1063/5.0245684},
    eprint = {https://pubs.aip.org/aip/jap/article-pdf/doi/10.1063/5.0245684/20382996/050901_1_5.0245684.pdf},
}

@article{Ohern,
  title = {Jamming at zero temperature and zero applied stress: The epitome of disorder},
  author = {O'Hern, Corey S. and Silbert, Leonardo E. and Liu, Andrea J. and Nagel, Sidney R.},
  journal = {Phys. Rev. E},
  volume = {68},
  issue = {1},
  pages = {011306},
  numpages = {19},
  year = {2003},
  publisher = {American Physical Society},
  doi = {10.1103/PhysRevE.68.011306},
  url = {https://link.aps.org/doi/10.1103/PhysRevE.68.011306}
}

@article{widmercooper2004,
  author = {Widmer-Cooper, Asaph and Harrowell, Peter and Fynewever, Hendrik},
  title = {How Reproducible Are Dynamic Heterogeneities in a Supercooled Liquid?},
  journal = {Phys. Rev. Lett.},
  volume = {93},
  pages = {135701},
  year = {2004},
  doi = {10.1103/PhysRevLett.93.135701}
}

@article{widmercooper2006,
  author = {Widmer-Cooper, Asaph and Harrowell, Peter},
  title = {Predicting the Long-Time Dynamic Heterogeneity in a Supercooled Liquid on the Basis of Short-Time Heterogeneities},
  journal = {Phys. Rev. Lett.},
  volume = {96},
  pages = {185701},
  year = {2006},
  doi = {10.1103/PhysRevLett.96.185701}
}

@article{widmercooper2008,
  author = {Widmer-Cooper, Asaph and Perry, Heather and Harrowell, Peter and Reichman, David R.},
  title = {Irreversible Reorganization in a Supercooled Liquid Originates from Localized Soft Modes},
  journal = {Nat. Phys.},
  volume = {4},
  pages = {711--715},
  year = {2008},
  doi = {10.1038/nphys1025}
}

@article{schoenholz2014,
  author = {Schoenholz, Samuel S. and Cubuk, E. D. and Sussman, Daniel M. and Kaxiras, Efthimios and Liu, Andrea J.},
  title = {A Structural Approach to Relaxation in Glassy Liquids},
  journal = {Nat. Phys.},
  volume = {12},
  pages = {469--471},
  year = {2016},
  doi = {10.1038/nphys3644}
}

@article{Volovik2019,
  author  = {G. E. Volovik and J. Rysti and J. T. M{\"a}kinen and V. B. Eltsov},
  title   = {Spin, Orbital, Weyl and Other Glasses in Topological Superfluids},
  journal = {Journal of Low Temperature Physics},
  year    = {2019},
  volume  = {196},
  number  = {1--2},
  pages   = {82--101},
  doi     = {10.1007/s10909-018-02132-z}
}

@article{Dzyaloshinskii1980,
  author  = {I. E. Dzyaloshinskii and G. E. Volovik},
  title   = {Poisson Brackets in Condensed Matter Physics},
  journal = {Annals of Physics},
  year    = {1980},
  volume  = {125},
  number  = {1},
  pages   = {67--97},
  doi     = {10.1016/0003-4916(80)90088-9}
}

@article{cubuk2016,
  author = {Cubuk, Ekin D. and Schoenholz, Samuel S. and Kaxiras, Efthimios and Liu, Andrea J.},
  title = {Structural Properties of Defects in Glassy Liquids},
  journal = {J. Phys. Chem. B},
  volume = {120},
  number = {26},
  pages = {6139--6146},
  year = {2016},
  doi = {10.1021/acs.jpcb.6b02144}
}

@article{cubuk2017,
  author = {Cubuk, E.D., and et al.,},
  title = {Structure-property relationships from universal signatures of plasticity in disordered solids},
  journal = {Science},
  volume = {358},
  number = {6361},
  pages = {1033--1037},
  year = {2017},
  doi = {10.1126/science.aai8830}
}

@article{richard2020,
  author = {Richard, D., and et al.},
  title = {Predicting Plasticity in Disordered Solids From Structural Indicators},
  journal = {Phys. Rev. Mater.},
  volume = {4},
  pages = {113609},
  year = {2020},
  doi = {10.1103/PhysRevMaterials.4.113609}
}

@article{Gartner,
  title = {Nonlinear plastic modes in disordered solids},
  author = {Gartner, Luka and Lerner, Edan},
  journal = {Phys. Rev. E},
  volume = {93},
  issue = {1},
  pages = {011001(R)},
  numpages = {5},
  year = {2016},
  publisher = {American Physical Society},
  doi = {10.1103/PhysRevE.93.011001},
  url = {https://link.aps.org/doi/10.1103/PhysRevE.93.011001}
}

@incollection{kleman2003book,
  author    = {Maurice Kleman and Oleg D. Lavrentovich},
  title     = {Topological Theory of Defects},
  booktitle = {Soft Matter Physics: An Introduction},
  publisher = {Springer},
  address   = {New York, NY},
  year      = {2003},
  pages      = {434--471},
  chapter    = {12},
  isbn       = {978-0-387-21759-8},
  doi        = {10.1007/978-0-387-21759-8_12},
  url        = {https://doi.org/10.1007/978-0-387-21759-8_12}
}

@article{onsager1949,
  author = {Onsager, Lars},
  title = {Statistical Hydrodynamics},
  journal = {Nuovo. Cim. (Suppl 2)},
  volume = {6},
  pages = {279--287},
  year = {1949},
  doi = {10.1007/BF02780991}
}

@article{kosterlitz1973,
  author = {Kosterlitz, J. M. and Thouless, D. J.},
  title = {Ordering, Metastability and Phase Transitions in Two-Dimensional Systems},
  journal = {J. Phys. C: Solid State Phys.},
  volume = {6},
  number = {7},
  pages = {1181--1203},
  year = {1973},
  doi = {10.1088/0022-3719/6/7/010}
}

@article{kosterlitz1974,
  author = {Kosterlitz, J. M.},
  title = {The Critical Properties of the Two-Dimensional XY Model},
  journal = {J. Phys. C: Solid State Phys.},
  volume = {7},
  number = {6},
  pages = {1046--1060},
  year = {1974},
  doi = {10.1088/0022-3719/7/6/005}
}

@article{mermin1979,
  author = {Mermin, N. David},
  title = {The Topological Theory of Defects in Ordered Media},
  journal = {Rev. Mod. Phys.},
  volume = {51},
  number = {3},
  pages = {591--648},
  year = {1979},
  doi = {10.1103/RevModPhys.51.591}
}

@book{chaikin1995,
  author = {Chaikin, Paul M. and Lubensky, Tom C.},
  title = {Principles of Condensed Matter Physics},
  publisher = {Cambridge University Press},
  address = {Cambridge},
  year = {1995},
  doi = {10.1017/CBO9780511813467}
}

@article{feynman1955,
  author = {Feynman, Richard P.},
  title = {Chapter II: Application of Quantum Mechanics to Liquid Helium},
  journal = {Prog. Low Temp. Phys.},
  volume = {1},
  pages = {17--53},
  year = {1955},
  publisher = {North-Holland},
  doi = {10.1016/S0079-6417(08)60077-3}
}

@book{kuramoto1984,
  author = {Kuramoto, Yoshiki},
  title = {Chemical Oscillations, Waves, and Turbulence},
  publisher = {Springer},
  address = {Berlin},
  year = {1984},
  doi = {10.1007/978-3-642-69689-3}
}

@article{shankar2022,
  author = {Shankar, Suraj and Souslov, Anton and Bowick, Mark J. and Marchetti, M. Cristina and Vitelli, Vincenzo},
  title = {Topological Active Matter},
  journal = {Nat. Rev. Phys.},
  volume = {4},
  pages = {380--398},
  year = {2022},
  doi = {10.1038/s42254-022-00445-3}
}

@article{Ronhovde2011,
  author  = {P. Ronhovde and Z. Nussinov},
  title   = {Detecting Hidden Spatial and Spatio-Temporal Structures in Glasses and Complex Physical Systems by Multiresolution Network Clustering},
  journal = {European Physical Journal E},
  year    = {2011},
  volume  = {34},
  pages   = {105},
  doi     = {10.1140/epje/i2011-11105-9}
}

@incollection{Nussinov2016,
  author    = {Z. Nussinov and P. Ronhovde and Dandan Hu and S. Chakrabarty and M. Sahu and Bo Sun and N. A. Mauro and K. K. Sahu},
  title     = {Inference of Hidden Structures in Complex Physical Systems by Multi-Scale Clustering},
  booktitle = {Information Science for Materials Discovery and Design},
  editor    = {Turab Lookman and Frank Alexander and Krishna Rajan},
  series    = {Springer Series in Materials Science},
  volume    = {225},
  publisher = {Springer},
  address   = {Cham},
  year      = {2016},
  chapter   = {6},
  pages     = {165--212},
  isbn      = {978-3-319-23870-8},
  doi       = {10.1007/978-3-319-23871-5_6}
}

@article{Ronhovde2012,
  author  = {P. Ronhovde and D. Hu and Z. Nussinov},
  title   = {Detection of Hidden Structures on All Scales in Amorphous Materials and Complex Physical Systems: Basic Notions and Applications to Networks, Lattice Systems, and Glasses},
  journal = {Scientific Reports},
  year    = {2012},
  volume  = {2},
  pages   = {329},
  doi     = {10.1038/srep00329}
}

@article{desmarchelier2024,
  author = {Desmarchelier, P. and Fajardo, S. and Falk, Michael L.},
  title = {Topological characterization of rearrangements in amorphous solids},
  journal = {Phys. Rev. E},
  volume = {109},
  pages = {L053002},
  year = {2024},
  doi = {10.1103/PhysRevE.109.L053002}
}

@article{baggioli2021,
  title = {Plasticity in Amorphous Solids Is Mediated by Topological Defects in the Displacement Field},
  author = {Baggioli, M. and Kriuchevskyi, I. and Sirk, T. W. and Zaccone, A.},
  journal = {Phys. Rev. Lett.},
  volume = {127},
  issue = {1},
  pages = {015501},
  numpages = {6},
  year = {2021},
  publisher = {American Physical Society},
  doi = {10.1103/PhysRevLett.127.015501}
}

@article{bera2025burgersrings,
  title         = {Burgers rings as topological signatures of Eshelby-like plastic events in glasses},
  author        = {Bera, Arabinda and Regev, Ido and Zaccone, Alessio and Baggioli, Matteo},
  journal       = {arXiv preprint arXiv:2505.23069},
  year          = {2025},
  eprint        = {2505.23069},
  archivePrefix = {arXiv},
  doi = {10.48550/arXiv.2505.23069},
}

@article{bera2025hedgehog,
  title     = {Hedgehog topological defects in 3D amorphous solids},
  author    = {Bera, Arabinda and Zaccone, Alessio and Baggioli, Matteo},
  journal   = {Nat. Commun.},
  volume    = {16},
  pages     = {5990},
  year      = {2025},
  doi       = {10.1038/s41467-025-61103-7}
}

@article{bera2024pnasnexus,
  title   = {Clustering of negative topological charges precedes plastic failure in 3D glasses},
  author  = {Bera, Arabinda and Baggioli, Matteo and Petersen, Timothy C. and Sirk, Timothy W. and Liu, Amelia C. Y. and Zaccone, Alessio},
  journal = {PNAS Nexus},
  year    = {2024},
  doi     = {10.1093/pnasnexus/pgae315}
}

@article{bera2025shearbands,
  title         = {Microscopic origin of shear bands in 2D amorphous solids from topological defects},
  author        = {Bera, Arabinda and Majumdar, Debjyoti and Sirk, Timothy W. and Regev, Ido and Zaccone, Alessio},
  journal       = {arXiv preprint arXiv:2507.09250},
  year          = {2025},
  eprint        = {2507.09250},
  archivePrefix = {arXiv},
  doi ={10.48550/arXiv.2507.09250},
}

@article{wu2023,
author={Wu, Z. W.
and Chen, Y.
and Wang, W.-H.
and Kob, W.
and Xu, L.},
title={Topology of vibrational modes predicts plastic events in glasses},
journal={Nat. Commun.},
year={2023},
volume={14},
number={1},
pages={2955},
doi ={10.1038/s41467-023-38547-w},
}

@article{wu2026geometry,
  author = {Wu, Z. W. and Barrat, J.-L. and Kob, W.},
  title = {On the Geometry of Topological Defects in Glasses},
  journal = {Nat. Commun.},
  volume = {17},
  pages = {217},
  year = {2026},
  doi = {10.1038/s41467-025-66923-1},
}

@article{wang2025topological,
  author = {Wang, Xin and Shang, Jin and Wang, Yujie and Zhang, Jie and Baggioli, Matteo},
  title = {Topological Defects Govern Plasticity and Shear Band Formation in Two-Dimensional Amorphous Solids},
  journal = {arXiv preprint arXiv:2507.03771},
  year = {2025},
  doi = {10.48550/arXiv.2507.03771},
}

@article{baggioli2026,
  author = {Baggioli, Matteo and Falk, Michael L. and Kob, Walter},
  title = {Topological Defects in Amorphous Solids},
  journal = {arXiv preprint arXiv:2604.07061},
  year = {2026},
  doi = {10.48550/arXiv.2604.07061}
}

@article{vaibhav2025,
  author = {Vaibhav, Vinay and Bera, Arabinda and Liu, Amelia C. Y. and Baggioli, Matteo and Keim, Peter and Zaccone, Alessio},
  title = {Experimental Identification of Topological Defects in 2D Colloidal Glass},
  journal = {Nat. Commun.},
  volume = {16},
  number = {1},
  pages = {55},
  year = {2025},
  doi = {10.1038/s41467-024-54857-z}
}

@article{schall2007,
  author = {Schall, Peter and Weitz, David A. and Spaepen, Frans},
  title = {Structural Rearrangements That Govern Flow in Colloidal Glasses},
  journal = {Science},
  volume = {318},
  number = {5858},
  pages = {1895--1899},
  year = {2007},
  doi = {10.1126/science.1149308}
}

@article{chikkadi2011,
  author = {Chikkadi, Vijay and Wegdam, Gerard and Bonn, Daniel and Nienhuis, Bernard and Schall, Peter},
  title = {Long-Range Strain Correlations in Sheared Colloidal Glasses},
  journal = {Phys. Rev. Lett.},
  volume = {107},
  number = {19},
  pages = {198303},
  year = {2011},
  doi = {10.1103/PhysRevLett.107.198303}
}

@article{ghosh2017,
  author = {Ghosh, Antina and Budrikis, Zoe and Chikkadi, Vijayakumar and Sellerio, Alessandro L. and Zapperi, Stefano and Schall, Peter},
  title = {Direct Observation of Percolation in the Yielding Transition of Colloidal Glasses},
  journal = {Phys. Rev. Lett.},
  volume = {118},
  number = {14},
  pages = {148001},
  year = {2017},
  doi = {10.1103/PhysRevLett.118.148001}
}

@article{chikkadi2014,
  author = {Chikkadi, V. and Miedema, D. M. and Dang, M. T. and Nienhuis, B. and Schall, P.},
  title = {Shear Banding of Colloidal Glasses: Observation of a Dynamic First-Order Transition},
  journal = {Phys. Rev. Lett.},
  volume = {113},
  number = {20},
  pages = {208301},
  year = {2014},
  doi = {10.1103/PhysRevLett.113.208301}
}

@article{liu2026,
  author = {Liu, Amelia C. Y. and Pham, Huyen and Bera, Arabinda and Petersen, Timothy C. and Sirk, Timothy W. and Mudie, Stephen T. and Tabor, Rico F. and Nunez-Iglesias, Juan and Zaccone, Alessio and Baggioli, Matteo},
  title = {Geometric Indicators of Local Plasticity in Glasses Measured by Scanning Small-Beam Diffraction},
  journal = {Acta Cryst. A},
  volume = {82},
  number = {1},
  pages = {4--17},
  year = {2026},
  doi = {10.1107/S2053273325009775}
}

@article{rosner2024,
  author = {R{\"o}sner, Harald and Bera, Arabinda and Zaccone, Alessio},
  title = {Unveiling the Asymmetry in Density within the Shear Bands of Metallic Glasses},
  journal = {Phys. Rev. B},
  volume = {110},
  number = {1},
  pages = {014107},
  year = {2024},
  doi = {10.1103/PhysRevB.110.014107}
}

@article{klix2015,
  title = {Discontinuous Shear Modulus Determines the Glass Transition Temperature},
  author = {Klix, Christian L. and Maret, Georg and Keim, Peter},
  journal = {Phys. Rev. X},
  volume = {5},
  issue = {4},
  pages = {041033},
  year = {2015},
  publisher = {American Physical Society},
  doi = {10.1103/PhysRevX.5.041033},
  url = {https://link.aps.org/doi/10.1103/PhysRevX.5.041033}
}

@article{smessaert2014,
  title = {Structural Relaxation in Glassy Polymers Predicted by Soft Modes: A Quantitative Analysis},
  author = {Smessaert, Anton and Rottler, J{\"o}rg},
  journal = {Soft Matter},
  volume = {10},
  issue = {43},
  pages = {8533--8541},
  year = {2014},
  publisher = {Royal Society of Chemistry},
  doi = {10.1039/C4SM01844B},
  url = {https://doi.org/10.1039/C4SM01844B}
}

@article{sahoo2026,
  title = {Topological defects govern plastic deformation and relaxation in 2D colloidal glasses driven by an optical vortex},
  author = {Sahu, Ratimanasee and Bera, Arabinda and Skanda, Rhutwik and Paul, Diptabrata and Schall, Peter and Kumar, G. V. P. and Zaccone, Alessio and Chikkadi, Vijayakumar},
  journal = {in communication},
  volume = {},
  issue = {},
  pages = {},
  year = {2026},
  publisher = {},
  doi = {},
  url = {}
}

@article{Keim17,
  title={Mermin--Wagner fluctuations in 2D amorphous solids},
  author={Illing, Bernd and Fritschi, Sebastian and Kaiser, Herbert and Klix, Christian L and Maret, Georg and Keim, Peter},
  journal={Proc. Natl. Acad. Sci. U.S.A.},
  volume={114},
  number={8},
  pages={1856--1861},
  year={2017},
  doi={https://doi.org/10.1073/pnas.1500763112},
  publisher={National Academy of Sciences}
}

@article{Royall13,
  title={Identification of structure in condensed matter with the topological cluster classification},
  author={Malins, Alex and Williams, Stephen R and Eggers, Jens and Royall, C Patrick},
  journal={The Journal of chemical physics},
  volume={139},
  number={23},
  year={2013},
  publisher={AIP Publishing}
}

\end{document}